\documentclass[epj]{svjour}
\usepackage{graphicx} 
\usepackage{epsfig,ulem,color} 
\usepackage{dcolumn}
\usepackage{bm}

\newcommand{\be}{\begin{eqnarray}}
\newcommand{\ee}{\end{eqnarray}}

\newcommand{\Sec}[1]{Sec.\@ \ref{#1}}
\newcommand{\Ref}[1]{Ref.\@ \cite{#1}}
\newcommand{\Fig}[1]{Fig.\@ \ref{#1}}

\begin{document}
\title{Imprint of the symmetry energy on the inner crust and strangeness content of neutron stars}
\author{Constan\c ca Provid\^encia\inst{1} \and Sidney S. Avancini
  \inst{2}\and Rafael Cavagnoli \inst{3} \and Silvia Chiacchiera \inst{1} \and Camille Ducoin\inst{4}
  \and Fabrizio Grill \inst{1} \and
  J\'er\^ome Margueron \inst{4} \and D\'ebora P. Menezes \inst{2} \and Aziz Rabhi
  \inst{1,5} \and Isaac Vida\~na \inst{1}}
\institute{Centro de F\'{\i}sica Computacional, Department of Physics, University of
  Coimbra, P3004-516 Coimbra, Portugal,
 \and 
Departamento de F\'{\i}sica, CFM, Universidade Federal de Santa Catarina, 
Florian\'opolis - SC - CP. 476, CEP 88.040 - 900, Brazil,
\and
Departamento de F\'{i}sica, IFM, Universidade Federal de Pelotas, Pelotas/SC, CP 354, CEP 96001-970, Brazil
\and
Institut de Physique Nucl\'eaire { de Lyon, CNRS/IN2P3, Universit\'e Claude Bernard Lyon 1, F-69622 Villeurbanne, France}
\and
Laboratoire de Physique de la Mati\`ere Condens\'ee, Facult\'e des Sciences de Tunis, Campus Universitaire, Le Belv\'ed\`ere-1060, Tunisia,
}
\date{Received: date / Revised version: date}

\abstract{
{ In this work we study the effect of the symmetry energy on several properties of neutron stars.
First, we discuss its effect on the density, proton fraction and pressure of the neutron star
 crust-core transition. We show that whereas the first two quantities present a clear correlation
with the slope parameter $L$ of the symmetry energy, no satisfactory correlation is seen between
the transition pressure and $L$. However, a linear combination of the slope and curvature 
parameters at $\rho=0.1$ fm$^{-3}$ is well correlated with the transition pressure.  In the second
part we analyze the effect of the symmetry energy on the pasta phase. It is shown that the size of
the pasta clusters, number of nucleons and the cluster proton fraction depend on the density 
dependence of the symmetry energy: a small $L$ gives rise to larger clusters. The influence of the
equation of state at subsaturation densities on the extension of the inner crust of the neutron star is 
also discussed. Finally, the effect of the effect of the density dependence of the symmetry energy on
the strangeness content of neutron stars is studied in the last part of the work. It is found that 
charged (neutral) hyperons appear at smaller (larger) densities for smaller values of the slope parameter
$L$. A linear correlation between the radius and the strangeness content of a star with a fixed mass is also
found.}
} 

\authorrunning{C. Provid\^encia {\it et al.,}}
\titlerunning{Imprint of the symmetry energy on the inner crust and strangeness content of neutron stars}

\maketitle
%


\section{Introduction}
\label{intro}
Isospin asymmetric nuclear matter is present in nuclei, especially in
those far away from the stability line, and in astrophysical systems,
particularly in neutron stars. Therefore, a well-grounded
understanding of the properties of isospin-rich nuclear matter is a
necessary ingredient for the advancement of both nuclear physics and
astrophysics \cite{baran05,li08,steiner05}. 
{ Nevertheless, in spite of the experimental \cite{lynch11} and theoretical \cite{baoanli11}
efforts carried out to study the properties of isospin-asymmetric  nuclear
systems, some of these properties are not well
constrained yet.} In particular, the density dependence of the symmetry
energy is still an important source of uncertainties. In this
work, we study how the density dependence of the symmetry energy
affects the equation of state (EOS) of asymmetric nuclear matter.
We will focus on three different problems. 

In the first part the correlations of the slope and curvature
parameters of the symmetry energy with the density, proton fraction and
pressure at the neutron star  crust-core transition are
analyzed.  
{The analysis is done with the microscopic Brueckner--Hartree--Fock (BHF) approach
and several phenomenological Skyrme and relativistic mean field (RMF) models to describe
the nuclear EOS \cite{vidana09}.}
A generalized liquid drop model (GLDM) based on a density
development around a reference density is introduced to allow the
identification of possible existing correlations between the
crust-core transition and  a limited set  of the coefficients of this
development \cite{ducoin10,ducoin11}. We show that the transition density and
the transition proton fraction are correlated with the symmetry energy
slope parameter $L$ and that the transition pressure shows no clear correlation with the symmetry
energy slope  at saturation. Nevertheless, a correlation exists between the
transition pressure and a linear combination of the symmetry
energy slope and curvature defined at $\rho=0.1$ fm$^{-3}$.

In the second part
the effect of the density dependence of  the symmetry energy  on the pasta
phase is discussed. It is shown that the number of
nucleons in the clusters, the cluster proton fraction and the size of the 
Wigner Seitz cell are very sensitive to the  
density dependence of the symmetry energy, and that the symmetry energy slope parameter $L$ may have quite dramatic effects
on the cell structure if it is very large or  small \cite{avancini2010,pasta3,grill12}. Rod-like and
slab-like pasta clusters have been obtained in all models except
one, with a large slope parameter $L$.  The effect of $L$
on the extension of the inner crust is also discussed. In particular, it
is shown that a smaller $L$ favors a wider slab phase. This phase
may allow the propagation of low frequency modes that will affect the
specific heat in a non-negligible way \cite{luc2011}. 

In the last part,
the effect of the density dependence of  the symmetry energy  on the
strangeness content of a neutron star is studied. The study is done with RMF models.
It is shown that there is still lacking information on the
nucleonic equation of state at supra-saturation densities and, in particular, 
on the hyperon interactions in nuclear matter. Therefore, the role
of exotic degrees of freedom on the interior of compact stars is still an open 
issue \cite{rafael11,panda12,providencia13a,providencia13}.

Finally, we show that some star properties are affected in a similar way by the density 
dependence of the symmetry energy and the hyperon content of the star. A linear 
correlation between the radius and the strangeness content of a star with a fixed mass is 
obtained \cite{providencia13}.

  In the following, before considering each one of these issues, we briefly review 
the main features of the asymmetric nuclear matter (ANM) equation of state and the different 
models considered in this work.


\section{Asymmetric Nuclear Matter EOS}
Assuming charge symmetry of the nuclear forces, the energy per particle of 
ANM can be expanded on the isospin asymmetry 
parameter, $\delta=(N-Z)/(N+Z)=(\rho_n-\rho_p)/\rho$, around the values of 
symmetric nuclear matter ($\delta=0$) in terms of even powers of $\delta$ as 
\begin{equation}
{E}(\rho,\delta)\simeq E_{SNM}(\rho)+E_{sym}(\rho)\delta^2  \ ,
\label{ea}
\end{equation}
where $E_{SNM}(\rho)$ is the energy per particle of symmetric matter and
$E_{sym}(\rho) =\frac{1}{2}\frac{\partial^2E}{\partial \delta^2}\Big
|_{\delta=0} $ is the symmetry energy.

It is common to characterize the density dependence of the symmetry energy around the
saturation density $\rho_0$ in terms of a few bulk parameters by expanding it in a Taylor series 
\begin{equation}
E_{sym}(\rho)=J+L\left(\frac{\rho-\rho_0}{3\rho_0}\right)
+\frac{K_{sym}}{2}\left(\frac{\rho-\rho_0}{3\rho_0}\right)^2
+ {\cal O}(3) \ ,
\label{s2c}
\end{equation}
where $J$ is the value of the symmetry energy at saturation and the
quantities $L$ and $K_{sym}$  are related 
to its slope  and curvature, respectively, at such density,
\begin{equation}
\begin{array}{c}
\displaystyle{L=3\rho_0\frac{\partial E_{sym}(\rho)}{\partial \rho} \Big |_{\rho=\rho_0}} \ , 
\displaystyle{K_{sym}=9\rho_0^2\frac{\partial^2 E_{sym}(\rho)}{\partial \rho^2} \Big |_{\rho=\rho_0}} .
\end{array}
\label{cs2b}
\end{equation}

\subsection{The BHF approach of ANM}
\label{sec:bhf}
The BHF approach of ANM \cite{bombaci91} starts with the construction 
of all the $G$ matrices which describe the effective interaction between two nucleons in the presence of 
a surrounding medium.  They are obtained by solving the well known Bethe--Goldstone equation , schematically
written as
\begin{eqnarray}
\label{bg}
G_{\tau_1\tau_2;\tau_3\tau_4}(\omega) &=&
V_{\tau_1\tau_2;\tau_3\tau_4}
+\sum_{ij}V_{\tau_1\tau_2;\tau_i\tau_j} \nonumber \\
&\times&\frac{Q_{\tau_i\tau_j}}{\omega-\epsilon_i-\epsilon_j+i\eta}
G_{\tau_i\tau_j;\tau_3\tau_4}(\omega)  
\end{eqnarray}
where $\tau=n,p$ indicates the isospin projection of the two nucleons in the initial, intermediate and
final states, $V$ denotes the bare NN interaction, 
$Q_{\tau_i\tau_j}$ the Pauli operator that allows only intermediate states compatible with the Pauli principle, 
and $\omega$, the so-called starting energy, corresponds to the sum of 
non-relativistic energies of the interacting nucleons. The single-particle energy 
$\epsilon_\tau$ of a nucleon with momentum $\vec k$ is given by
\begin{equation}
\epsilon_{\tau}(\vec k)=\frac{\hbar^2k^2}{2m_{\tau}}+Re[U_{\tau}(\vec k)] \ ,
\label{spe}
\end{equation}
where the single-particle potential $U_{\tau}(\vec k)$ represents the mean field ``felt'' by a nucleon due to its
interaction with the other nucleons of the medium. In the BHF approximation, $U(\vec k)$ 
is calculated through the ``on-shell energy'' $G$-matrix, and is given by
\begin{eqnarray}
\label{spp}
U_{\tau}(\vec k)&=&
\sum_{\tau'}
\sum_{ |\vec{k}'| <  k_{F_{\tau'}}} \\
&&\langle \vec{k}\vec{k}'
\mid
G_{\tau\tau';\tau\tau'}(\omega=\epsilon_{\tau}(k)+\epsilon_{\tau'}(k'))
\mid \vec{k}\vec{k}' \rangle_A \nonumber
\end{eqnarray}
where the sum runs over all neutron and proton occupied states and where the matrix elements are properly 
antisymmetrized. We note here that the so-called continuous prescription has been adopted for the single-particle
potential when solving the Bethe--Goldstone equation \cite{jekeunne76}. Once a self-consistent 
solution of Eqs.\ (\ref{bg})--(\ref{spp}) is achieved, the energy per particle can be calculated as
\begin{equation}
{E}(\rho,\delta)=\frac{1}{A}\sum_{\tau}\sum_{|\vec{k}| <  k_{F_{\tau}}}
\left(\frac{\hbar^2k^2}{2m_{\tau}}+\frac{1}{2}Re[U_{\tau}(\vec k)] \right) \ .
\label{bea}
\end{equation}

The BHF calculation carried out in this work uses the realistic Argonne V18 \cite{wiringa95} nucleon-nucleon 
interaction supplemented with a nucleon three-body force of Urbana type which, for the use in BHF calculations, 
was reduced to a two-body density dependent force by averaging over the spatial, spin and isospin coordinates of the third nucleon
in the medium \cite{baldo99}. 

\subsection{Phenomenological models}

Phenomenological approaches, either relativistic or nonrelativistic, are based
on effective interactions that are frequently built to reproduce properties of nuclei.
Skyrme interactions \cite{skyrme} and RMF models \cite{rmf} are
among the most commonly used ones. They are briefly described in the next.

\subsubsection{Skyrme interaction}
The standard form of a Skyrme interaction reads 
\begin{eqnarray}
V(\bf r_1,\bf r_2)&=&t_0\left(1+x_0P^\sigma\right)\delta(\bf r) \nonumber \\
&+&\frac{1}{2}t_1\left(1+x_1P^\sigma\right)\left({\bf k'}^2\delta(\bf r) + \delta(\bf r) {\bf k}^2 \right) \nonumber \\
&+&t_2\left(1+x_2P^\sigma\right){\bf k'}\cdot \delta(\bf r){\bf k} \nonumber \\
&+&\frac{1}{6}t_3\left(1+x_3P^\sigma\right)\left(\rho(\bf R)\right)^\alpha \delta(\bf r) \nonumber \\
&+&iW_0\left(\sigma_1+\sigma_2\right)\left({\bf k'}\times\delta(\bf r){\bf k} \right) \ ,
\label{skyrme}
\end{eqnarray}
where ${\bf r=r_1-r_1}, {\bf R=(r_1+r_2)/2}$, ${\bf k}=(\nabla_1-\nabla_2)/2i$ is the relative momentum
acting on the right, ${\bf k'}$ its conjugate acting on the left and $P^\sigma=(1+\bf \sigma_1\cdot \bf\sigma_2)/2$ 
is the spin exchange operator. The last term, proportional to $W_0$, corresponds to the zero-range spin-orbit term.
It does not contribute in homogeneous systems and thus will be ignored for the rest of this article. 

Most of these forces are, by construction, well behaved close to the saturation density and moderate isospin 
asymmetries. Nevertheless, only certain combinations of their parameters are well determined empirically. Consequently, 
there is a proliferation of different Skyrme interactions that produce a similar EOS for symmetric nuclear
matter but predict a very different one for pure neutron matter. A few years
ago, Stone {\it et al.,} \cite{stone03} tested extensively and systematically the 
capabilities of almost 90 existing Skyrme parametrizations to provide good 
neutron star candidates. They found that only 27 of these parametrizations passed the
restricted tests they imposed, the key property being the density dependence of 
the symmetry energy. These forces are SLy0-SLy10 \cite{chabanat95} and SLy230a \cite{chabanat97} 
of the Lyon group,  SkI1-SkI5 \cite{ski1-5} and SkI6 \cite{ski6} of the SkI family, Rs and Gs \cite{rsgs}, 
SGI \cite{sgi}, SkMP \cite{skmp}, SkO and SkO' \cite{sko}, SkT4 and SkT5 \cite{skt}, and the early SV \cite{sv}.
The results for the Skyrme forces shown in this work have been obtained  with these 27 forces and 
the additional parametrizations SGII \cite{sgi}, RATP \cite{ratp}, SLy230b \cite{chabanat97}, NRAPR \cite{steiner05}, 
LNS \cite{lns}, BSk14 \cite{bsk14}, BSk16 \cite{bsk16} and BSk17 \cite{bsk17}. 
We should mention, however, that more stringent constraints to the Skyrme forces have been very recently 
presented by Dutra {\it et al.,} in Ref.\ \cite{jirina}. These authors have examined the suitability of 240 Skyrme
interactions with respect to eleven macroscopic constraints derived mainly from experimental data and the empirical 
properties of symmetric nuclear matter at and close to saturation. They have found that only 5 of the 240 forces 
analyzed satisfy all the constraints imposed. We note that among the parametrizations used in this work, only the NRAPR and LNS  
ones belong to this restricted set.

\subsubsection{RMF models}
RMF models are based on effective Lagrangian densities where nucleons interact with and through an
isoscalar-scalar field $\sigma$, an isoscalar-vector field $\omega^{\mu}$, an isovector-vector field 
$\boldsymbol{\rho}^{\mu}$, and an isovector-scalar field $\boldsymbol{\delta}$. In this work we 
consider models with constant couplings 
and non-linear terms~\cite{bb}, and with density dependent couplings~\cite{tw}. Within the first class 
of models, that we will designate by Non Linear Walecka Models (NLWM), we consider
NL3~\cite{nl3} and GM1, GM3~\cite{gm1} with non linear $\sigma$ terms, NL3$_{\omega\rho}$
including also non-linear $\omega\rho$ terms that allow the
modulation of the density dependence of the symmetry
energy~\cite{hor01}, TM1 ~\cite{tm1}  with non linear $\sigma$ and
$\omega$ terms, FSU~\cite{fsu} and IU-FSU~\cite{iufsu} with
non-linear $\sigma$, $\omega$ and $\omega\rho$ terms. 
{The last} two
parametrizations were constrained by the collective response of
nuclei to the isoscalar monopole giant resonance (ISGMR) and the
isovector dipole giant resonance (IVGDR).
Within the second class of models with density dependent couplings
we consider TW~\cite{tw}, DD-ME2~\cite{ddme2} and DD-ME$\delta$~\cite{roca2011}:
DD-ME2, as all the non-linear parametrizations considered, does not
include the $\delta$ meson, and was adjusted to experimental data
based on finite nuclei properties; DD-ME$\delta$ contains the
$\delta$ meson and was fitted to microscopic ab-initio
calculations in nuclear matter and finite nuclei properties. Both
models present similar properties for the symmetry energy, however,
DD-ME2 has a larger incompressibility at saturation.

The Lagrangian density for these models typically reads
\begin{equation}
\mathcal{L}=\sum_{i=p,n}\mathcal{L}_{i}\mathcal{\,+L}_{{\sigma }}%
\mathcal{+L}_{{\omega }}\mathcal{+L}_{{\rho
}}\mathcal{+L}_{{\delta}}
\mathcal{+L}_{{nl}}, 
\label{lagdelta}
\end{equation}
where the nucleon Lagrangian is
\begin{equation}
\mathcal{L}_{i}=\bar{\psi}_{i}\left[ \gamma _{\mu }iD^{\mu }-M^{*}\right]
\psi _{i}  
\label{lagnucl},
\end{equation}
with
\begin{eqnarray}
iD^{\mu } &=&i\partial ^{\mu }-\Gamma_{\omega}\Omega^{\mu }-\frac{\Gamma_{\rho }}{2}{\boldsymbol{\tau}}%
\cdot \boldsymbol{\rho}^{\mu }, \label{Dmu} \\
M^{*} &=&M-\Gamma_{\sigma}\sigma-\Gamma_{\delta
}{\boldsymbol{\tau}}\cdot \boldsymbol{\delta}, \label{Mstar}
\end{eqnarray}
and the meson  Lagrangian densities are given by
\begin{eqnarray}
\mathcal{L}_{{\sigma }} &=&\frac{1}{2}\left( \partial _{\mu }\sigma \partial %
^{\mu }\sigma -m_{\sigma}^{2}\sigma ^{2}\right)   \\
\mathcal{L}_{{\omega }} &=&\frac{1}{2} \left(-\frac{1}{2} \Omega _{\mu \nu }
\Omega ^{\mu \nu }+ m_{\omega}^{2}\omega_{\mu }\omega^{\mu } \right)  \\
\mathcal{L}_{{\rho }} &=&\frac{1}{2} \left(-\frac{1}{2}
\mathbf{R}_{\mu \nu }\cdot \mathbf{R}^{\mu
\nu }+ m_{\rho }^{2}\boldsymbol{\rho}_{\mu }\cdot \boldsymbol{\rho}^{\mu } \right) \\
\mathcal{L}_{ {\delta }} &=&\frac{1}{2}(\partial _{\mu }\boldsymbol{\delta}%
\partial ^{\mu }\boldsymbol{\delta}-m_{\delta }^{2}{\boldsymbol{\delta}}^{2}) \\
\mathcal{L}_{{nl}} &=&-\frac{1}{3!}\kappa \sigma ^{3}-\frac{1}{4!}%
\lambda \sigma ^{4}+\frac{1}{4!}\xi \Gamma_{\omega}^{4}(\omega_{\mu}\omega^{\mu })^{2} \nonumber \\
&+&\Lambda_\omega \Gamma_\omega^2 \Gamma_\rho^2 \omega_{\mu
}\omega^{\mu } \boldsymbol{\rho}_{\mu }\cdot
\boldsymbol{\rho}^{\mu }~.
\end{eqnarray}

In the above equations $\Gamma_i$ ($i=\sigma, \omega, \rho, \delta$) denote, depending on the model, the constant or 
density-dependent coupling parameters. Finally, we note that the photon and electron Lagrangian densities
\begin{eqnarray}
\mathcal{L}_{{\gamma }} &=&-\frac{1}{4}F _{\mu \nu }F^{\mu\nu }  \\
\mathcal{L}_{e}&=&\bar{\psi}_{e}\left[ \gamma _{\mu }\left(i\partial^{\mu }+eA^{\mu}\right)-m_e \right]
\psi _{e}  
\label{lage},
\end{eqnarray}
and the term $- e (1+\tau_3)A^{\mu}/2$ should be added to Eqs.\ (\ref{lagdelta}) and (\ref{Dmu}), respectively,
when describing non-homogeneous matter. 
$\beta$-equilibrium matter requires the inclusion of the lepton
Lagrangian density only.

\subsection{The Generalized Liquid-Drop Model (GLDM)}\label{sec:1a}
The liquid-drop model equation of state is based on a density expansion around
the saturation point, its main features  being: the saturation density
$\rho_0$, the energy per nucleon at saturation $E_0$, the
incompressibility coefficient $K_0$ which characterizes the curvature
of the EOS, and the isovector coefficients, namely the symmetry energy $J$, its slope $L$
and the symmetry incompressibility $K_{\rm sym}$. We focus on the bulk-matter
equation of state, and
no surface effects are considered. In order to discuss the link that can be
drawn between laboratory constraints and EOS properties in situations quite
away from the experimental data, in particular, at high density or at low proton fraction, we introduce a ``generalized liquid-drop model" (GLDM) 
\cite{ducoin10,ducoin11}, still addressing the bulk EOS, which expresses the EOS as an expansion of arbitrary order around a chosen reference 
density $\rho_{\rm ref}$ (not necessarily the saturation density $\rho_0$): 
\be
\label{EQ:GLDM_EOS}
E_{\rm GLDM}(\rho,\delta)&=&\sum_{n=0}^{\mathcal N}\left(c_{{\rm IS}, n}+c_{{\rm IV}, n}\,\delta^2\right)\frac{x^n}{n!}\nonumber\\
&+&(E_{\rm kin}-E_{\rm kin}^{\rm para})\,, 
\ee
with $x={(\rho-\rho_{\rm ref})}/{3 \rho_{\rm ref}}$.
 The coefficients $c_{{\rm IS},n}$ and $c_{{\rm IV},n}$, where the index `IS'
 (`IV') stands for isoscalar (isovector), are  associated with the derivatives 
of the energy $E(\rho,\delta=0)$ and of the symmetry energy $E_{sym}(\rho)$
\be
c_{{\rm IS},n}(\rho_{\rm ref}) &=& (3\rho_{\rm ref})^n \frac{\partial^n E}{\partial\rho^n}(\rho_{\rm ref},0) \nonumber\\
c_{{\rm IV},n}(\rho_{\rm ref}) &=& (3\rho_{\rm ref})^n \frac{\partial^n
  E_{sym}}{\partial\rho^n}(\rho_{\rm ref}).
\nonumber
\ee
For $\rho_{\rm ref}=\rho_0$, the lower-order coefficients are usual nuclear matter properties:
$c_{{\rm IS},0}=E_0$ (saturation energy), 
$c_{{\rm IS},2}=K_0$ (incompressibility), 
$c_{{\rm IS},3}=Q_0$ (skewness),
$c_{{\rm IV},0}=J$ (symmetry energy),
$c_{{\rm IV},1}=L$ (symmetry-energy slope),
$c_{{\rm IV},2}=K_{\rm sym}$ (symmetry incompressibility),
$c_{{\rm IV},3}=Q_{\rm sym}$.
Since the effective laboratory constraints are rather related to subsaturation density, the coefficients of a GLDM with reference density $\rho_{\rm ref}<\rho_0$ are expected to be better constrained than the standard saturation coefficients. Also, the neutron-star core-crust transition properties are better correlated with coefficients defined at subsaturation density because the reference point is closer to the transition point.

The isovector channel of the  EOS in Eq. (\ref{EQ:GLDM_EOS})  has a parabolic contribution  accounted for by the isovector coefficients multiplied by $\delta^2$, and a minimal extra-parabolic 
correction,  the model-independent kinetic term $E_{\rm kin}-E_{\rm kin}^{\rm
  para}$, 
\be 
E_{\rm kin}
&=&\frac{(3\pi^2/2)^{5/3}}{10 m\pi^2}\rho^{2/3}\left[ (1+\delta)^{5/3} + (1-\delta)^{5/3} \right] ,\nonumber\\
E^{\rm para}_{\rm kin}&=&\frac{(3\pi^2/2)^{5/3}}{10 m\pi^2}\rho^{2/3}\left[2+\frac{10}{9}\delta^2 \right]\, ,\nonumber
\ee 
 that introduces the divergence of  energy-density curvature in the proton-density direction at small proton density  and, therefore, avoids that the spinodal
contour reaches pure neutron matter. In the above equation $m$ refers to  the nucleon mass.

The EOS obtained with any nuclear model can be associated with its corresponding GLDM.
In the limit of an infinite expansion (GLDM$_{\infty}$), the symmetric matter EOS $E(\rho,0)$ and the symmetry energy $E_{sym}(\rho)$ are exactly equivalent to the complete model EOS, the only remaining difference, for the EOS of asymmetric matter, being the extra-parabolic terms of the interaction part. 
In the following, we will address the correlations that could be found between GLDM coefficients and the neutron-star core-crust transition properties, 
being focused, in particular, on the isovector coefficients.


\section{Crust-core transition}
\label{sec:1}

The EOS of nuclear matter can be constrained by laboratory data and
astrophysical observations.  It has been shown in Refs.\ \cite{ducoin10} and \cite{ducoin11} that an accurate 
determination of  the symmetry energy and its
slope and curvature at a subsaturation density, $\rho=0.1$ fm$^{-3}$, allows
a quite accurate prediction  of the core-crust transition properties.

{Nuclear} models could then be used to
restrict the range of the core-crust transition properties in
neutron stars and contribute to the interpretation
of astrophysical observations. Pulsar glitches are an example \cite{link99}, 
since the transition pressure is an essential input
to infer the neutron-star mass-radius relation from glitch
observations.

In the present section we will show how some of the EOS properties are
correlated with the crust-core transition properties.

\begin{figure}[tbh]
\begin{center}
\begin{tabular}{ccc}
\includegraphics[width =.7\linewidth]{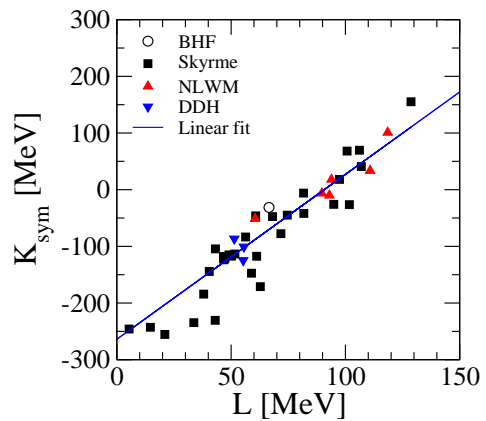}
\end{tabular}
\end{center}
\caption{(Color online)
Correlation between  $K_{\rm sym}$  and  $L$.
}
\label{Fig:asyEOS}
\end{figure}

\subsection{Correlation between $L$ and $K_{\rm sym}$}\label{sec:1b}
When different nuclear models are compared, the GLDM coefficients taken at
saturation density present some correlations between them. This is a
manifestation of the fact that the effective constraints from laboratory data
do not fix the nuclear properties at saturation, but rather at a lower
density. Indeed, the nuclei whose properties are used to fit the effective
nuclear density functionals have a mean density that is lower than $\rho_0$.
Many different nuclear models tend to converge to a symmetry energy close to 
25 MeV at $\rho \simeq$ 0.12 fm$^{-3}$. A similar convergence can be observed 
for the symmetry energy slope, whose value becomes close to 100 MeV at $\rho \simeq$ 0.06 fm$^{-3}$
(see Refs. \cite{ducoin10} and \cite{ducoin11} for a more detailed analysis of these features). 

These convergent trends away from the saturation point imply some correlations
in the expansion coefficients around saturation density. An example is the
existing correlation between $J$ and $L$ \cite{ducoin10,ducoin11}: since
the  symmetry energy is better constrained at subsaturation
density, 
higher values of $L$ have to be compensated by higher values of $J$.
Another example is the  correlation
between $L$ and $K_{\rm sym}$   shown in
Fig. \ref{Fig:asyEOS}. The remarkably strong $L$-$K_{\rm
  sym}$ correlation plays an important role in the links that can be drawn between the GLDM coefficients (namely, laboratory constraints) and the core-crust transition pressure in neutron star, a sensitive input for the interpretation of pulsar glitches.

\subsection{Correlation between $L$ and the core-crust transition properties}
\label{sec:1c}

New experimental perspectives for the measurement of $L$ have drawn interest in trying to correlate this quantity to the core-crust transition properties. Such correlations have to be considered with care, taking into account that fake relations may appear when the study is limited to a restricted nuclear model or family of models, with internal correlations that disappear if different kinds of functional models are considered. Reliable correlations between GLDM coefficients and the core-crust transition properties have to remain true independently of the difference between nuclear models, as long as these models account for the existing experimental constraints.

Studying the correlations between $L$ and the core-crust transition properties
in the framework of various models (effective relativistic and Skyrme models,
and BHF calculations), we arrive to the conclusion that: (i) $L$ is well
correlated with the transition density and proton fraction ($\rho_{\rm t}$,
$Y_{\rm p,t}$) and (ii) $L$ is not satisfactorily correlated with the
transition pressure $P_{\rm t}$. We have denoted the proton fraction by
$Y_p=\rho_p/\rho$, and used the subscript $t$ to refer to the properties at
the crust-core transition. We summarize below the explanation of this situation (see Refs. \cite{ducoin10,ducoin11} for a more detailed analysis).

\begin{figure}[tbh]
\begin{center}
\begin{tabular}{ccc}
\includegraphics[width =0.9\linewidth]{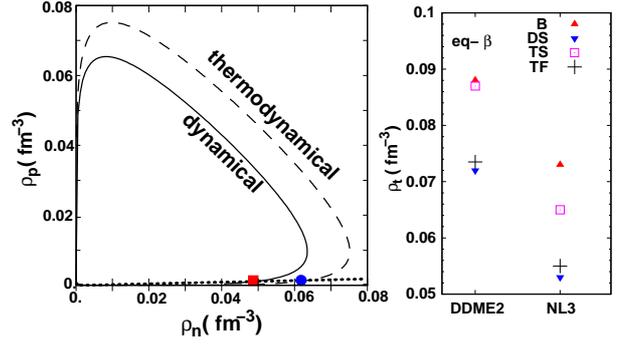}&  
\end{tabular}
\end{center}
\caption{(Color online) {Left panel: Comparison between the thermodynamical (dashed line) and dynamical (full line) spinodals. The dotted line
represents the $\beta$-equilibrium EOS and the red square and blue dot define the crust-core
transition within, respectively, the dynamical and thermodynamical
spinodal. Right panel: comparison of the transition density obtained from
different approaches (binodal, dynamical spinodal, thermodynamical spinodal
and Thomas Fermi calculation)  for two RMF models: DDME2 and NL3.
}}
\label{Fig:spin}
\end{figure}

\begin{figure*}[tbh]
\begin{center}
\begin{tabular}{ccc}
\includegraphics[width =0.8\linewidth]{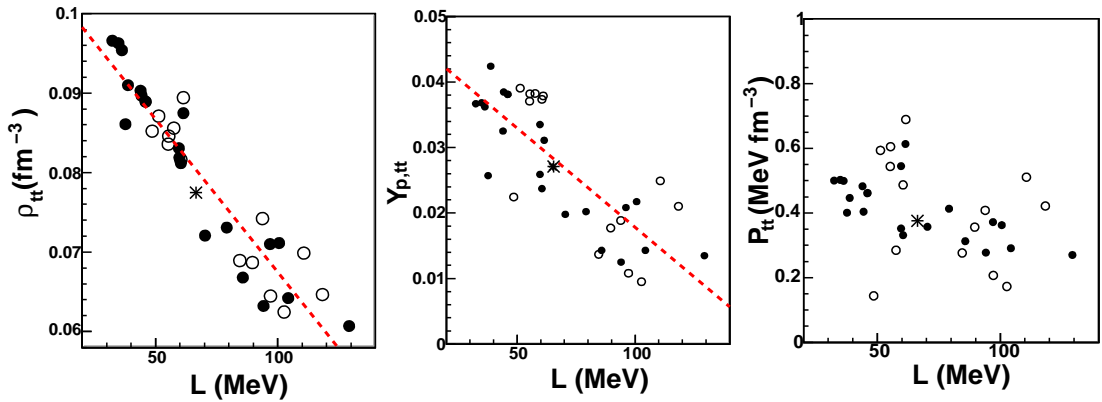}
\end{tabular}
\end{center}
\caption{(Color online)
$L$ versus a) the total transition density $\rho_{tt}$ (left panel),
b)
the proton fraction at the transition $Y_{p,tt}$ (middle panel), and
c) transition pressure $P_{tt}$ (right panel) for
different nuclear models: Skyrme (full symbols), relativistic models
(empty symbols), BHF (asterisk).}
\label{Fig:L-rhot-Ypt}
\end{figure*}

\subsubsection{Core-crust transition: thermodynamical versus dynamical calculations}

We will next discuss how to determine the core-crust transition point in neutron stars.
Cold neutron-star matter is in $\beta$-equilibrium, and is
transparent to neutrinos,  thus, for a given
nuclear density, the proton fraction of homogeneous nuclear matter is essentially
determined by the symmetry energy at this density. The core-crust border is
the transition from the homogeneous matter of the core to the clusterized
matter of the inner crust. 
For the very neutron-rich matter of a cold
neutron star, this transition is well approximated by  the
dynamical spinodal border \cite{dynamical}, as shown in Refs.\ \cite{avancini2010,grill12} where it was
compared with pasta phase calculations. The dynamical spinodal is the
density region where the homogeneous nuclear matter is unstable against
finite-size density fluctuations, 
eventually leading to cluster formation, and takes into account both finite
size effects and the Coulomb interaction.

In fact,  it is expected that the transition density lies in the metastable region
between the binodal surface and the dynamical spinodal surface. The binodal surface is 
defined  in the $\rho,\, Y_p,\, T$ phase space as the surface where the gas and liquid phases coexist, 
and corresponds to an  upper limit for the extension of the pasta phase because it does not take into 
account neither Coulomb nor finite size effects.

A  simplified definition of  the transition point is determined by the crossing
between the $\beta$-equilibrium line and the thermodynamical spinodal border
\cite{avancini06,camille08,vidana08,xu09,pasta1,pasta2}. The thermodynamical spinodal,
 the bulk liquid-gas instability region in nuclear matter,
 touches the binodal surface at the critical point, which, for a given temperature,
occurs at a density and
proton fraction close to the crust-core transition density and  proton
fraction. 
This
transition point is sensitively different from the real transition point,
since the thermodynamical spinodal region is larger than the dynamical one, as
can be seen in Fig. \ref{Fig:spin}.  
{The bulk (thermodynamical) instability of nuclear matter is at the
  origin of the instability against clusterization that affects star
  matter. This second instability region, named dynamical spinodal, 
takes into account surface terms and the Coulomb interaction \cite{pethick95},
 both leading to a reduction of the dynamical spinodal 
with respect to the bulk one.}
As a result, the transition point calculated on the basis of the thermodynamical
spinodal (that will be denoted by the index $tt$) is at a significantly higher
density than the actual transition point, well approximated on the basis of
the dynamical spinodal (this transition point will be denoted by the index
$td$). However, we have verified that the properties of both transition points
are well correlated \cite{ducoin11}. Since the $tt$ point allows to study more directly the link between the GLDM coefficients and the transition properties, we start by discussing this thermodynamical transition point, keeping in mind that it represents a shifted version of the core-crust transition. The correlation effects observed in the case of the $tt$ transition are expected to apply as well to the more realistic $td$ transition point.

\subsubsection{Correlations between $L$ and the transition density ($\rho_{t}, Y_{p,t}$)}

The correlation between $L$ and the transition density point (total density and proton fraction) is quite robust, 
as can be seen in Fig. \ref{Fig:L-rhot-Ypt} (left and middle panels).
This results from two effects that reinforce each-other:
a)
a larger value of $L$ means a smaller symmetry energy at subsaturation
density, i.e. a more neutron-rich $\beta$-equilibrium (lower proton fraction
$Y_{p,t}$). According to the shape of the spinodal, this also means a lower
density $\rho_{t}$; b) the value of $L$ also has an impact on the spinodal border: a larger $L$ is associated with a spinodal border at lower density.
To explain this second effect, let us consider the energy-density curvature of
neutron matter, taken at the symmetric matter spinodal density $\rho_{\rm s}$. This particular density is chosen in order to cancel the isoscalar terms, and to concentrate on the isovector ones. Thus, this quantity reads:
\be
\label{Eq:Cnms}
C_{\rm NM,s}
&=&
\frac{2}{3\rho_0}L + 
\frac{1}{3\rho_0}\sum_{n\geq 2}
 c_{{\rm IV},n}
\frac{x_{\rm s}^{n-2}}{(n-2)!}
\left[ \frac{n+1}{n-1}x_{\rm s}+\frac{1}{3}  \right]\nonumber\\
&+&\frac{\partial^2 \left[\rho(E_{\rm kin}-E_{\rm kin}^{\rm para})
    \right]}{\partial \rho^2}
\ee
with $x_{\rm s}=(\rho_{\rm s}-\rho_0)/(3\rho_0)$.
The leading term is proportional to $L$, and the following only has a quite
weak effect 
(see Ref. \cite{ducoin11}).

\subsubsection{Lack of correlation between $L$ and the transition pressure $P_{t}$}

In the case of the transition pressure $P_{t}$, no satisfactory correlation with $L$ emerges when different kinds of models are involved, 
as can be seen on Fig. \ref{Fig:L-rhot-Ypt} (right panel).
To understand this result, it is useful to express $P$ as a development in terms of the GLDM coefficients :

\be
P(\rho,\delta)
&=&\frac{\rho^2}{3\rho_0}\left[ L \delta^2 
+ \sum_{n\geq 2}\left(c_{{\rm IS}, n}+c_{{\rm IV}, n}\delta^2\right)\frac{x^{n-1}}{(n-1)!}\right]\nonumber\\
&+& \rho^2 \frac{\partial (E_{\rm kin}-E_{\rm kin}^{\rm para})}{\partial \rho}
\, . 
\label{eq:pressure}
\ee
The lack of $L$-$P_{t}$ correlation that is observed results from three main effects, which are opposed and compensate each-other:
a) the leading term of the density development of the pressure is proportional to $L$, so $P_{t}$ should increase with $L$ ;
b) the transition density $\rho_{t}$ has been shown to decrease with $L$, and
the pressure should decrease if the density decreases;
c) the second term of the development, whose sign is negative, is proportional
to the symmetry incompressibility $K_{\rm sym}$, which is larger for larger
$L$. Effects b) and c) are opposite to a).

\begin{figure}[tbh]
\begin{center}
\begin{tabular}{ccc}
\includegraphics[width =0.65\linewidth]{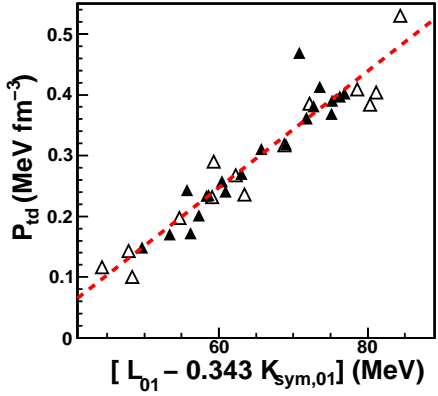}& 
\end{tabular}
\end{center}
\caption{(Color online)
 Correlation obtained between $P_{td}$ and a linear combination of
 $L_{0.1}$ and $K_{\rm sym,0.1}$ [see text,  Eq.~(\ref{eq:corrP-comb})].
}
\label{Fig:GLDM-Ptd}
\end{figure}

This conclusion has been reached by analyzing the different contributions to the link between $L$ and $P_{\rm t}$ through a variational study of Eq. (\ref{eq:pressure}). Two kinds of variations were considered: (i) variations of the transition point (density and proton fraction) and (ii) variations of the GLDM coefficients; both kinds of variations are correlated with $L$. 

Thus, for a given $L$ value, the transition pressure obtained will essentially
depend on the nuclear model that is used. Different models may predict either
an increasing or decreasing correlation of $P_{\rm t}$ with $L$ \cite{ducoin11}, and indeed opposite predictions exist in the literature \cite{xu09,moustakidis09}.

\subsection{Correlations between transition pressure and GLDM coefficient combinations}

Although the link between the transition pressure and the GLDM coefficients is
quite delicate, as it has been shown in the previous section, it is important
to find a way to get a reliable estimation of $P_t$ in relation with
quantities that could be constrained by laboratory data. For this reason, we
looked beyond a simple correlation with $L$, which we did not obtain, and
investigated the role of other GLDM coefficients (see
Ref. \cite{ducoin11} for a detailed discussion). 
One of the most promising correlations involves a linear combination of
$L_{0.1}$ and $K_{\rm sym,0.1}$, denoting, respectively, the symmetry energy
slope and curvature taken at the reference density $\rho_{\rm ref}$=0.1
fm$^{-3}$ instead of $\rho_0$. 
We have considered the pressure at the crossing of the dynamical spinodal,
$P_{td}$, 
and
performed a linear fit with two variables:
\be
P_{td}(L_{01},K_{\rm sym,01})&=&a \times L_{01} + b \times K_{\rm sym,01} +c\,,
\ee 
The following relation was obtained
\be
P_{td}(L_{01},K_{\rm sym,01}) &=& 9.59\times 10^{-3} \times [L_{01} - 0.343
  \times K_{\rm sym,01}] \nonumber\\
&&- 0.328  \mbox{  MeV fm}^{-3}~.
\label{eq:corrP-comb}
\ee
This correlation is represented on Fig. \ref{Fig:GLDM-Ptd}, where it is
compared with corresponding $P_{td}$ versus $L$ plot.
A similar relation has been verified recently  by the authors of
Ref. \cite{Newton13}, although a different slope coefficient is obtained,
which they attribute to the different method used to determine the transition point.

\section{Symmetry energy and the pasta phase}
\label{sec:SEandPasta}
Neutron stars and proto-neutron stars are believed to have in the inner crust 
a special non-homogeneous matter known as
{\it pasta phase}.
The pasta phase is a frustrated system that arises from
the competition between the strong and the electromagnetic interactions
\cite{pethick,horo,maruyama,watanabe05,watanabe08} and appears
at densities of the order of 0.001 - 0.1 fm$^{-3}$ \cite{pasta1,watanabe08}
in neutral nuclear matter 
or in a smaller density range \cite{pasta2,bao}
in $\beta$-equilibrium stellar matter.
The basic shapes of these structures 
(droplets (bubbles), rods (tubes) and slabs for three, two and one dimensions
respectively) were first discussed in \cite{pethick}, where the authors 
joked on the resemblance with {\it lasagna}, {\it spaghetti} and so on, from 
which the phase name {was chosen}.
It was shown in \cite{oyama07} that  the EOS of the inner crust
is particularly sensitive to the density dependence of the symmetry energy,
and, therefore, it is expected that the pasta phase structure will depend on it.

In the following we discuss how the density dependence of the symmetry
energy affects the pasta phase and the inner-crust structure within a nuclear
relativistic mean-field approach. 
We adopt, in line with many authors, the following simplifying view: for some given 
conditions (temperature, density, proton fraction or chemical equilibrium) 
a single geometry will be the physical one. 
{ That is, in practice, we compute the free energy of homogeneous matter and 
the five structures and choose as the physical one that with the smaller free energy.}
 The denser regions (clusters) will form a regular lattice 
that we study in the Wigner-Seitz (WS) approximation.

First, we will discuss the
 pasta phase properties within a naive picture 
 {that} uses the Gibbs conditions of coexisting phases  and 
includes by hand the surface and Coulomb contributions \cite{pasta3,pasta1,pasta2}. This description will
be denoted by the coexisting phases
(CP) method and will allow the identification of the main pasta features that
depend on the symmetry energy.

Next, we will present a complete self-consistent 
calculation of the pasta phase within a relativistic 
mean field density dependent
Thomas-Fermi approach (TF)~\cite{pasta1}. 
What before was described as a two-density system with a sharp interface, is now 
described as a system with smoothly varying densities.
In \Ref{avancini2010} we compared the TF and CP approaches and found that 
the TF method confirms the main trends given by the more naive CP method, but predicts 
a wider and richer pasta phase.
{It is worth emphasizing}
that other approaches are available to build the pasta phase, each with its 
advantages and disadvantages. For example, a Hartree-Fock-Bogoliubov (HFB) calculation
would allow to include the shell effects, neglected in the RMF approach, however, 
within this framework only the spherical symmetry could be addressed. For a comparison of HFB and 
TF results in this context see \Ref{grill12}.

\begin{figure}[ht]
  \centering
\begin{tabular}{c}
\includegraphics[width=0.8\linewidth]{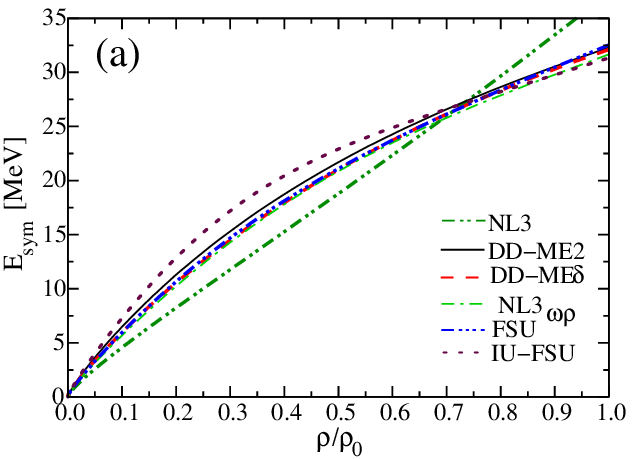}\\ 
\includegraphics[width=0.8\linewidth]{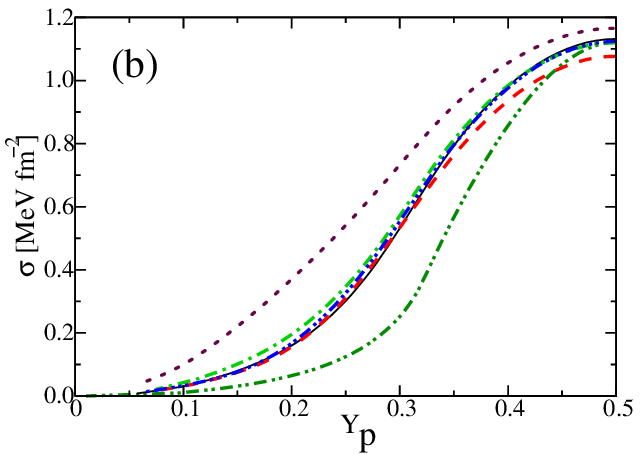}
\end{tabular}
\caption{(Color online) a) Symmetry energy vs density and b) surface tension at $T=0$ vs the proton
  fraction for NL3, NL3$\omega\rho$, FSU, IU-FSU, DD-ME2 and DD-ME$\delta$.}
 \label{EsymFig}
\end{figure}

Finally, we will use the TF EOS for the pasta phase as an ingredient to 
compute the star structure, and we will discuss how the presence of the pasta phase 
affects the inner crust.

\subsection{Surface tension and symmetry energy}

Before entering the discussion of the pasta phase, and in order to compare the models used in
this part, namely NL3, NL3$\omega\rho$, FSU, IU-FSU, DDME2 and DDME$\delta$,
we plot in Fig.~\ref{EsymFig} the symmetry energy vs density, and the surface tension 
$\sigma$ vs the proton fraction for all of them.
The surface tension coefficient $\sigma$ (shown here for $T=0$) was obtained in the TF approach along 
the lines explained in Ref.
\cite{pasta3} and it is used in a parametrized form 
as an input in our CP calculation. 
The properties of the pasta phase in the CP approach
depend crucially on $\sigma$;  also in the TF approach, 
{where surface terms are generated self-consistently, it}
is a useful guideline to interpret the results.

\begin{table}[tbh]
  \centering
  \begin{tabular}{lcccccc}
\hline
\hline 
 &  {\scriptsize NL3}     &  {\scriptsize NL3$\omega\rho$} & {\scriptsize FSU}
   & {\scriptsize IU-FSU}  & {\scriptsize DDME2}   & {\scriptsize DDME$\delta$} \\
\hline  
$\rho_0$
                     &  0.148    &  0.148          & 0.148  & 0.155   & 0.152  & 0.152 \\
$J$
&  37.3    &  31.7           & 32.6   & 31.3    & 32.3    &  32.4  \\
$L$
&  118.3   &  55.2           & 60.5   & 47.2    & 51.4    &  52.9\\
$K_0$
&  270.7   &     272.0       &  230.0 &  231.2  & 250.8   & 219.1 \\
\hline
\hline 
  \end{tabular}
\caption{Properties at saturation: density $\rho_0$ (fm$^{-3}$), symmetry
  energy value $J$ (MeV) and slope $L$ (MeV), 
and incompressibility $K_0$ (MeV), for the models discussed 
in \Sec{sec:SEandPasta}. 
}
\label{tab:JL}
\end{table}
 
To help the discussion of this section, we present in Table \ref{tab:JL} the isovector properties at saturation
predicted by the six models considered here. As shown in the table the six models predict very similar values 
for the symmetry energy at saturation, namely, $J$ varies between 31.3 and 32.6 MeV,  except for NL3 that 
has a quite high value,37.3 MeV. However, there is a larger dispersion of the symmetry
energy slope at saturation, $L$, with values that go from 47.2 MeV (IU-FSU), 
to 118.3 MeV for NL3.

The properties of the pasta will reflect these facts, with IU-FSU and
NL3 behaving in a quite different way, while all the other models
show similar results. The slope $L$ has a direct influence on
the surface tension and surface thickness of the clusters.
A smaller $L$ corresponds generally to a larger surface tension
and smaller neutron skin thickness~\cite{hor01}, as can be
seen by comparing the surface tensions of the above models.
The decrease of the surface tension  with the slope $L$ may be understood
from the fact that the neutron pressure at a density close to 0.1 fm$^{-3}$,
a typical density at the nucleon surface, is essentially proportional to the
slope $L$ \cite{vidana09,brown}.
Therefore, a larger value of $L$ will favor neutron drip and a smaller surface tension, {\it i.e.,} particles at the surface
are not so tightly bound to the nucleus.

\subsection{Pasta phase}

\subsubsection{Coexisting phases (CP) method}

As first approximation, the pasta phase is calculated within the CP method \cite{pasta1}. 
In this section we focus on nuclear matter with a fixed proton fraction 
$Y_p$ and impose charge neutrality by setting $\rho_e=\rho_p$. However, the same scheme 
could be applied to $\beta$-equilibrium stellar matter: in this case, the 
species fractions would be defined by the conditions of chemical equilibrium and charge neutrality.

As in \cite{pasta1,maruyama}, for a given total density $\rho$,
the pasta structures are built with different
geometrical forms in a background nucleon gas. This is achieved by calculating
from the Gibbs conditions $P^I=P^{II}$,  $\mu_i^I=\mu_i^{II}$ where I and II 
label the high and low density phase respectively,
the density and the proton fraction of the
pasta and of the background gas.
The density of electrons is considered uniform.
The total pressure and total energy density 
of the system are given, respectively, by $P=P^I+P_e$ and 
\begin{equation}
{\cal E}= f {\cal E}^I + (1-f) {\cal E}^{II} + {\cal E}_e 
+ {\cal E}_{surf} + {\cal E}_{Coul},
\label{totener}
\end{equation}
where $f$ is the volume fraction of phase I, the proton fraction 
{can be obtained from}
$$f \rho_p^I+ (1-f)\rho_p^{II}=Y_p\rho,$$
and  ${\cal E}_e$, ${\cal E}_{surf}$  and  ${\cal E}_{Coul}$
denote electron, surface and Coulomb energy densities.
By minimizing  ${\cal E}_{surf} + {\cal E}_{Coul}$ with respect
to the size of the droplet/bubble, rod/tube or slab we get
\cite{maruyama}
${\cal E}_{surf} = 2 {\cal E}_{Coul},$ and
\begin{equation}
{\cal E}_{Coul}=\frac{2 F}{4^{2/3}}(e^2 \pi \Phi)^{1/3}
\left(\sigma D (\rho_p^I-\rho_p^{II})\right)^{2/3},
\end{equation}
where $F=f$ for droplets and $F=1-f$ for bubbles,
 $\sigma$ is the surface energy coefficient,
$D$ is the dimension of the system and $\Phi$ is
the geometric factor:  
$$\Phi=
\left(\frac{2-D F^{1-2/D}}{D-2}+F \right) \frac{1}{D+2}, \, D=1,2,3.$$

In the following discussion the 
parameter sets for the models NL3, NL3$\omega\rho$, FSU and IU-FSU
will be considered and the effect  
of the symmetry energy on the pasta phase discussed.
One expects, generally speaking, two types of effects:
a smaller $L$ corresponds to a larger surface tension for asymmetric matter
\cite{pasta3} and a larger $J$ leads to a more 
isospin-symmetric liquid phase. In models with a larger surface tension the pasta phase sets in at higher 
densities and neutron drip is unfavored, giving 
rise to a lower density background gas.

\begin{figure}[t]
  \centering
\begin{tabular}{c}
\includegraphics[width=0.9\linewidth]{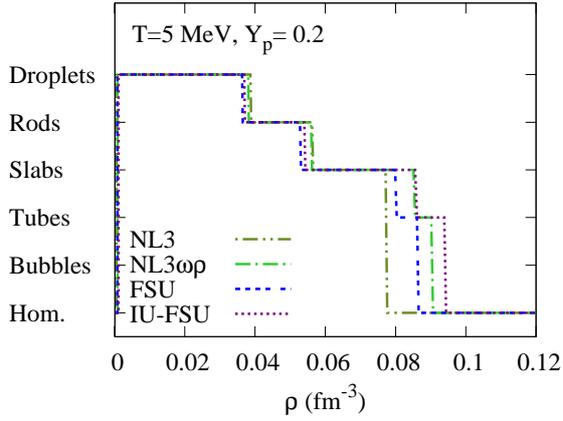}
\end{tabular}
\caption{(Color online) Pasta phases in the CP method for $T=5$ MeV and $Y_p=0.2$. 
  Results for NL3, NL3$\omega\rho$, FSU, IU-FSU.
}
\label{fig:ShapeCP}
\end{figure}

In Fig. \ref{fig:ShapeCP}
the range of the different pasta phases for matter
with a proton fraction $Y_p=0.2$ and temperature $T=5$ MeV is plotted for these
four models. Some comments are in order: for the proton
fraction considered no model presents the bubble configuration, and for NL3
the tube configuration is also missing; the onset of the rod and slab
configurations  are quite model independent, while the transition to the core
reflects the symmetry energy behavior, in particular, the transition density
is smaller for larger values of $L$, as discussed above.

\begin{figure*}[t]
  \centering
\begin{tabular}{ccc}
\includegraphics[height=4.7cm]{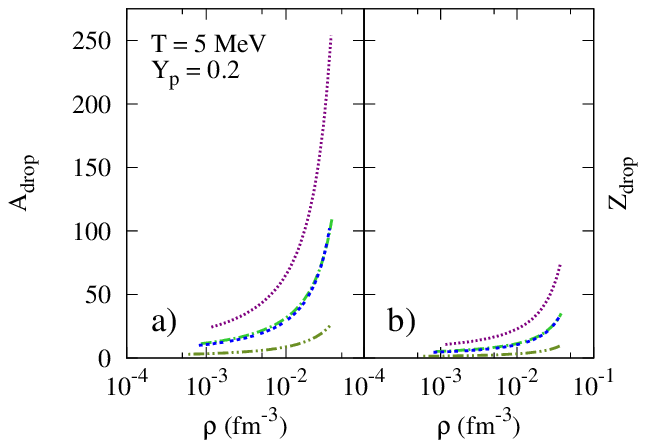}
\includegraphics[height=4.7cm]{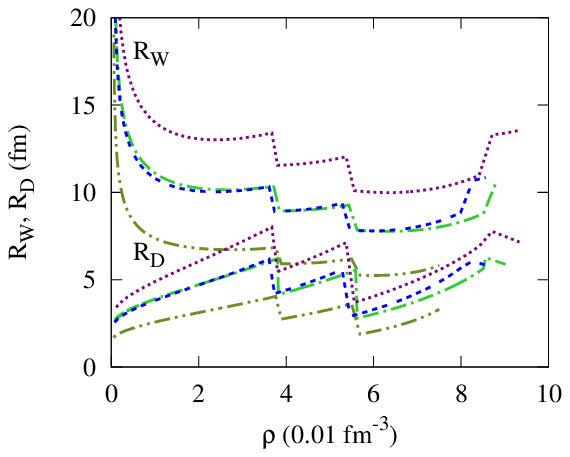}
\includegraphics[height=4.7cm]{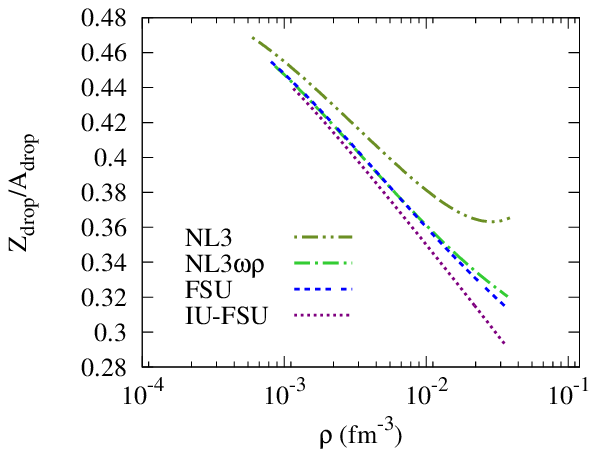}
\end{tabular}
\caption{(Color online) Properties of pasta obtained in the CP method,  at $T=5\,$MeV and $Y_{p}=0.2$. 
Results for NL3, NL3$\omega\rho$, FSU
  and IU-FSU parametrizations. Left: The number of a) nucleons $A$ and b) protons  $Z$ in a droplet. 
Middle: cluster and Wigner Seitz cell radius. Right: The ratio $Z/A$ in the droplet phase.
}
\label{fig:propCP}
\end{figure*}
In Fig. \ref{fig:propCP} 
we show some pasta properties, mainly for the droplet phase.
Let us denote by  $A_{\rm drop}$  the number
of nucleons belonging to a droplet (the type of structure that appears at the
lowest densities in the non-homogeneous phase) and by $Z_{\rm drop}$ its charge content. 
Notice that we adopt the prescription to define both quantities as excesses with 
respect to the background nucleon gas.
In the left panel
we display the results for a)  $A_{\rm drop}$ and
b) $Z_{\rm drop}$ for  the above parametrizations.
The onset of the droplet
phase is characterized by a discontinuity on the number of nucleons inside the
cluster: a minimum number of nucleons  is necessary  to compensate the
surface energy, which is larger for models with a smaller $L$. It should be
referred, however, that in a TF calculation, where the surface energy is
calculated self-consistently, this discontinuous
behavior will not occur. 
A change in  the isovector
channel of the model NL3 as in NL3$\omega\rho$ leads to a large effect on the
number of nucleons in the droplet, increasing this number to more than the
double. In fact, a smaller symmetry energy slope corresponds to a larger
  surface energy and neutrons do not drip out so easily. The
number of nucleons obtained within NL3$\omega\rho$  is consistent with the 
results of \cite{hempel10} within a statistical model. The other two
models, FSU and IU-FSU, also present
larger  nuclei than NL3, the heaviest ones corresponding to the model IU-FSU,
{which bears} the smallest slope $L$.

The radius of the Wigner Seitz cell together with the cluster 
radius is plotted in the middle panel of Fig. \ref{fig:propCP} 
as a function of density. 
The ordering of the radii obtained in the different parametrizations
reflects perfectly the ordering of their surface tensions \cite{pasta3},  that, 
in turn, is closely linked to the symmetry energy density dependence.
NL3 has by far the smallest surface energy at $Y_{p}=0.2$, while IU-FSU
has the largest: correspondingly, NL3 has the smallest  
Wigner Seitz cell and droplets and IU-FSU the largest ones.

In the right panel of Fig. \ref{fig:propCP} 
the ratio $Z_{\rm drop}/A_{\rm drop}$
  is plotted as a function of density. This ratio decreases with density and is model
  dependent. A decrease of the proton fraction of the clusters with density
  was also obtained in \cite{raduta10}. 
The models with a smaller symmetry
  energy slope have smaller proton fractions. A smaller slope implies that
  neutrons drip out of the cluster with more difficulty giving rise to
 neutron richer clusters.  
 Also,  a smaller slope 
corresponds to a smaller $ J$, and,  thus,
 a smaller $L$ 
  favors  less symmetric clusters.

In summary, within the coexistence method we have shown that  models with a smaller symmetry
energy slope have larger clusters with a smaller proton fraction and larger Wigner-Seitz cells. We will next discuss
the predictions of a Thomas Fermi calculation of the pasta phase
\cite{grill12}, which generally agree with the above conclusions.

\subsubsection{Thomas-Fermi (TF) approach}
\label{sec:pastatf}
In the Thomas-Fermi approximation of the
non-uniform {\it npe} matter, the fields are assumed to vary slowly so that
the baryons can be treated as moving in locally constant fields at each point \cite{maruyama,pasta1}.
We obtain  the finite temperature
semiclassical TF approximation
 based on the density functional formalism \cite{pasta3} and start from the grand canonical potential density:
\begin{equation}
 {\omega } = \omega ( \{f_{i+}\},\{f_{i-}\},\sigma_0,\omega_0,\rho_0,\delta_0) =
{\cal E}_t-T{\cal S}_t-\sum_{i=p,n,e}\mu_i \rho_i ~ ,~
\label{grand}
\end{equation}
where $\{f_{i+}\}$($\{f_{i-}\}$), $i=p,n,e$  stands for the  protons,
neutrons and electrons  positive (negative) energy distribution functions and
 ${\cal S}_t = {\cal S} +{\cal S}_e$ , ${\cal E}_t = {\cal E} +{\cal E}_e $ are
the total entropy  and energy densities
respectively \cite{avancini2010}. 
 The equations of motion for the meson fields (see Ref.~\cite{pasta1})  follow from the variational 
conditions:
\begin{equation}
\frac{\delta}{\delta \sigma_0(\mathbf r)} \Omega =
\frac{\delta}{\delta \omega_0(\mathbf r)} \Omega = 
\frac{\delta}{\delta \rho_0(\mathbf r)} \Omega  =
\frac{\delta}{\delta \delta_0(\mathbf r)} \Omega =  0 ~ ,
\nonumber \label{meson}
\end{equation}
where $\Omega = \int d^3 r ~ \omega$. 

The numerical algorithm for the description of the neutral $npe$ matter at finite temperature is
a generalization of the zero temperature case which
was discussed in detail in \cite{avancini2010,pasta1}.
The Poisson equation is always solved by using the appropriate Green
function according to the  spatial dimension of interest and the
Klein-Gordon equations are solved by expanding the meson fields in a harmonic
oscillator basis with one, two or three dimensions based on the method
presented in \cite{avancini2010,pasta1}.

\begin{figure}[thb]
  \centering
  \includegraphics[width=0.9\linewidth]{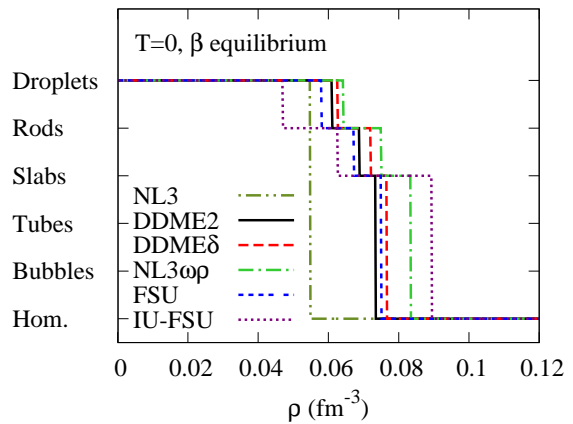}
  \caption{(Color online) Pasta phases in TF method at T=0 in $\beta$-equilibrium matter. 
    Results for NL3, NL3$\omega\rho$, FSU, IU-FSU, plus two density dependent
    hadronic parametrizations, DDME2 and DDME$\delta$.
     }
 \label{fig:ShapeTF}
\end{figure}

We next present results for the pasta phase of $\beta$-equili-\\brium matter obtained within a TF
calculation at T=0. 
Due to the $\beta$-equilibrium condition the proton fraction is very small and
only three different 
shapes occur: droplet, rod and slab.
The transition densities between the shapes
are compared in Fig. \ref{fig:ShapeTF} 
for the six models mentioned above.
All of the three shapes appear in the inner crust except for NL3, which only
predicts the existence of droplets. In fact, in~\cite{oyama07} it was shown
that models with a large $L$, like NL3,  would not predict the existence of 
non-droplet pasta
shapes in $\beta$-equilibrium matter.
As discussed before, the  slope $L$
defines the crust-core transition within models in the same framework. It is
seen, however, that although IU-FSU has the largest crust-core transition
density to the core, it also has the smallest transition density to the rod
and slab configurations. This behavior probably reflects the large surface
energy of IU-FSU that favors smaller surface to volume geometries.

In Fig. \ref{fig3} we show
the neutron density at the cell center and at  the cell border (left panel), the cluster proton
fraction at the cluster center (middle panel)
and the
total atomic number of a {cluster (right panel)
\cite{grill12}. 
As in the CP method, also here the background nucleon gas is subtracted 
when defining the cluster properties $Y_{\rm p, drop}$ and A. 
Notice that the number of nucleons belonging to a cluster or to a cell (A, N and Z), 
are univocally determined by the calculation only in the case of droplets. 
For the slab and rod phases, by construction, the problem is only solved in one or two 
dimensions. The values of A, Z and N for these shapes were obtained assuming 
representative sizes for the rod length and for the slab cross-section \cite{grill12}.

These results are coherent with the ones calculated
within the coexisting phases method.
The density of drip-\\ped neutrons is smallest
for IU-FSU, the model with the smallest slope $L$. 
Moreover, IU-FSU (NL3) has the largest (smallest)
number of nucleons in the clusters, corresponding to the smallest (largest)
slope $L$, and at the cluster center the proton fraction is largest for models
with the largest symmetry energy $J$. 

\begin{figure*}[th]
  \centering
\begin{tabular}{ccc}
\includegraphics[width=0.3\linewidth]{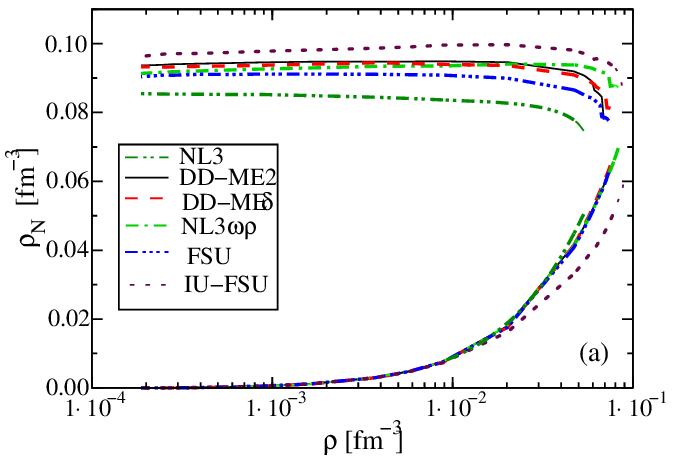}& 
\includegraphics[width=0.3\linewidth]{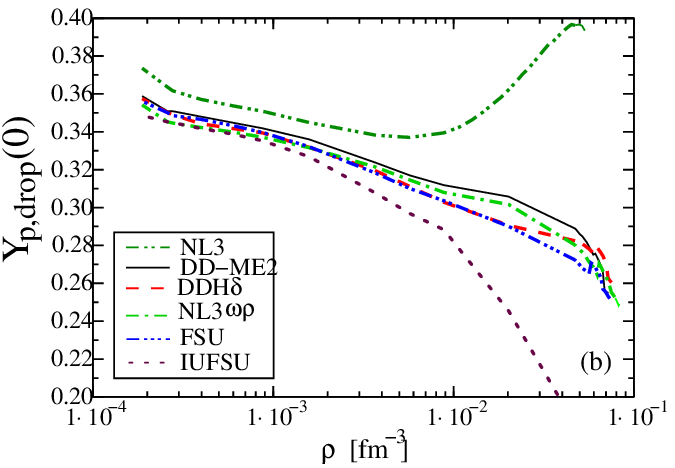}& 
\includegraphics[width=0.3\linewidth]{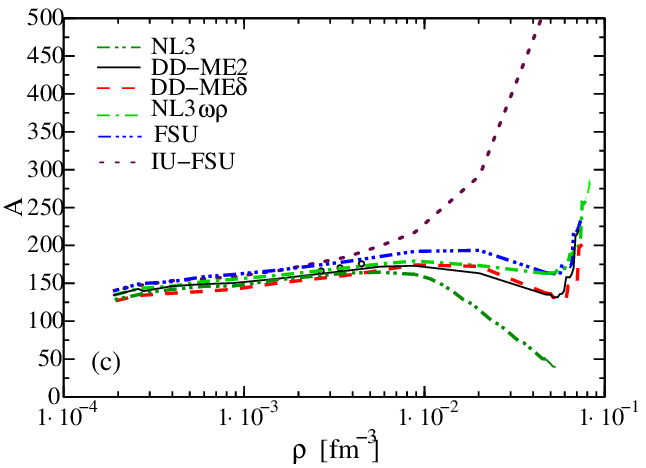}  
\end{tabular}
\caption{(Color online) Neutron density at the cell center (upper curves) and border (lower curves) (a),
    cluster proton fraction at the cluster center (b) and the  total atomic
    number of a {cluster } (c)
for the same models as in Fig.~\ref{fig:ShapeTF}.}
 \label{fig3}
\end{figure*}

\begin{center}
\begin{figure*}[thb]
  \centering
\begin{tabular}{ccc}
\includegraphics[width=0.3\linewidth]{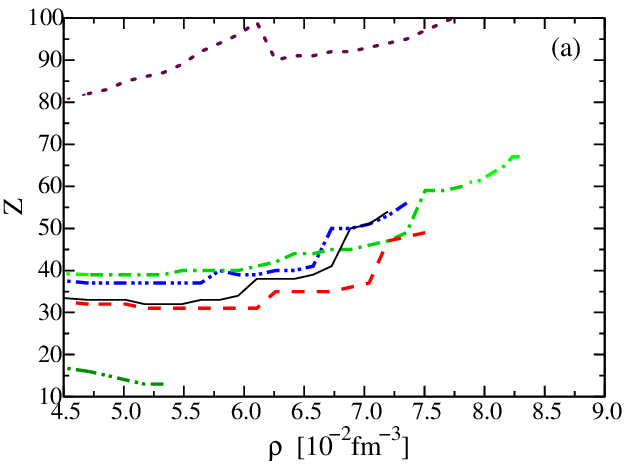}& 
\includegraphics[width=0.3\linewidth]{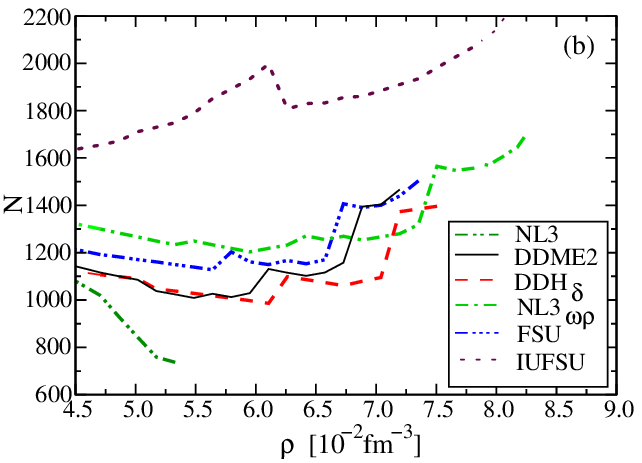}&  
\includegraphics[width=0.3\linewidth]{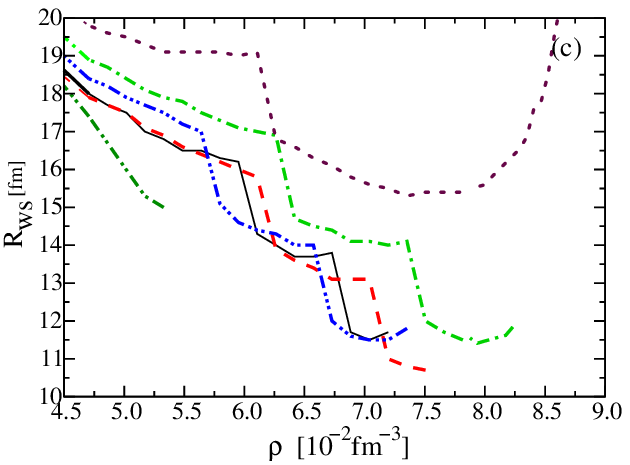} 
\end{tabular}
\caption{(Color online) a) Proton number,  b) neutron number  and c) radius of a WS cell in the
    pasta phase regions. Same models as in Fig.~\ref{fig:ShapeTF}.}
 \label{ZNRws}
\end{figure*}
\end{center}
{ In Fig.~\ref{ZNRws} we show some properties of the WS cells: the proton (a) and neutron (b) content of each cell (N and Z), 
and the cell radius (c).}

The properties of the models used are reflected on the cluster
structure. A small symmetry energy slope $L$ gives
rise to larger cells, with a larger proton and neutron number,
while the opposite occurs for a large $L$. Models with a similar
symmetry energy ($\sim31-32$ MeV) and  slope $L$ ($\sim 50-60$ MeV) at
saturation density  behave in a similar way, both in
the droplet phase and the non-droplet pasta phase regions. On the other hand,
models  like NL3, with a very large symmetry energy and slope $L$,
and  IU-FSU, with a quite small $L$, have quite different
behaviors. NL3 does not present any non-droplet pasta phase in the inner crust
of $\beta$-equilibrium matter, predicts the smallest proton
and neutron numbers and  the Wigner-Seitz radius 
{in} 
almost all the inner crust range of densities.

 All of the models, except NL3,
 predict the existence of  slab like configurations in $\beta$-equilibrium
matter. These {\it lasagna}-like structures may have an important contribution to
the specific heat of the crust ~\cite{luc2011}.

\begin{table}[tbh]
  \centering
  \begin{tabular}{lcccccccccccccc}
\hline
\hline 
 $M$ &    $\epsilon_c$ &   $R_{hs}$  &   $R_{sr}$ &    $R_{rd}$  &  $R_{dBPS}$ &   $R$\\
   ($M_\odot$) &    (fm$^{-4}$) & (km)  & (km)  & (km)  &  (km) &(km) \\
\hline
\\
&&&{\bf \scriptsize FSU}\\
  1.00  &    1.81 &  11.09 &  11.14 &  11.23 &  11.93&   12.80\\
  1.44  &    3.07 &  11.25 &  11.28&   11.33 &  11.75 &  12.28\\
  1.60  &    4.39&   10.92 &  10.94&   10.99 &  11.30&   11.66\\
 1.66 & 7.04& 10.27&10.29 &10.32 & 10.56 & 10.84\\
\\
&&&{\bf \scriptsize  IU-FSU}\\
 1.00 &   1.78&   10.82&   10.90&   10.96 &  11.64&   12.46\\
  1.44 &   2.61&   11.29&   11.34 &  11.38&   11.81&   12.30\\
  1.60 &   3.18 &  11.26&   11.31 &  11.34 &  11.69&   12.09\\
 1.80& 6.69 &10.48 &10.50&10.52&10.74&11.00\\
\\
&&&{\bf \scriptsize NL3} \\
 1.00  &    1.12&   12.57  &    - &   -&  13.38&   14.53\\
  1.44 &     1.39 &  13.34 &   - &   - &  13.89&   14.63\\
  1.60 &     1.49 &  13.53 &   - &   - &  14.01&   14.66\\
 2.78& 4.42 &12.78&-&-&13.12&13.29\\
\\
&&&{\bf \scriptsize  NL3$\omega\rho$}\\
 1.00  &  1.28 &  11.40 &  11.47&   11.59 &  12.46 &  13.42\\
  1.44 &   1.52 &  12.37 &  12.42 &  12.50 &  13.11&   13.75\\
  1.60 &   1.62 &  12.63&   12.67 &  12.75 &  13.28 &  13.84\\
2.68& 4.62&12.49&12.50&12.52&12.70&12.87\\
    \hline\hline
  \end{tabular}
  \caption{Central energy density, distance to the center of the star at the
  phase   transitions: homogeneous matter--slab phase ($ R_{hs}$),
slab phase--rod phase ($R_{sr}$), rod-phase--droplet phase  ($ R_{rd}$),
droplet phase--outer crust ($R_{dBPS}$) and radius of a 1.0, 1.4, and 1.6
$M_\odot$ star for several EOS, for different models. For each model, 
the maximum mass configuration is also shown. 
}\label{profile}
\end{table}

\subsection{Inner crust structure}

In this section we analyse how the EOS at subsaturation densities affects the
inner-crust extension.

The Tolman--Oppenheimer--Volkov (TOV) equations are solved  to determine the density
 profile of neutron stars with masses 1, 1.44, 1.6 $M_\odot$.  
These are stars with representative masses: the
lowest one is smaller than the smallest neutron stars detected until
now, 1.44 $M_\odot$ is the mass of the Hulse-Taylor pulsar, and the largest
mass is chosen to be smaller than the maximum mass described by FSU.
Besides these three values,  the TOV equations are also solved for the maximum mass  star. 
The stellar matter EOS's are  built
according to the following scheme \cite {grill13}: a) the EOS in the core 
is obtained including only nucleonic degrees of freedom, electrons and muons,
solving the equations for the meson fields in the mean-field
approximation and imposing both $\beta$-equilibrium and charge
neutrality; b) for the outer crust, the BPS  (Baym-Pethick-Sutherland)
EOS \cite{bps} is
considered; c) the inner crust, corresponding to the range of
densities between the neutron drip ($\sim 2 \times 10^{-4}$  fm$^{-3}$) and the crust-core transition, 
is obtained from the TF calculation of $\beta$-equilibrium non-homogeneous
matter \cite{pasta3,grill12,pasta1,pasta2}. 

In Table \ref{profile}  
we  display some of the features of the inner crust structure
  according to different models.
All models considered have a slab and a rod
phase which together define the non-droplet pasta extension,  except for
  NL3. For this model the inner crust is only formed by droplets in a neutron gas
background. 
For the sake of readability, some of the results given in Table 
\ref{profile} are plotted in Fig. \ref{crust}  and \ref{tov}.
In Fig.  \ref{crust}   the thickness of the crust
(full symbols) and inner crust (empty symbols) are given in the left
panel,  the thickness of the total non-droplet pasta phase (full symbols) and
the slab phase (empty symbols) are plotted in the middle panel, and
the fraction of the inner crust with respect to the total crust  is
given in the right panel. The different models are ordered
according to the magnitude of the slope $L$,  which increases from
left to right.
In Fig. \ref{tov} we represent instead
the crust profile, identifying the transition between the
different configurations  with marks (black lines and symbols).  In
the same figure,  it is also shown the crust profile calculated with an EOS obtained joining the BPS EOS directly
to the  homogeneous stellar matter EOS (red dashed lines). In this
last case the transition from the BPS to the homogeneous matter EOS is
shown by a red full point.


\begin{figure*}[thb]
\centering
  \includegraphics[width=0.9\linewidth]{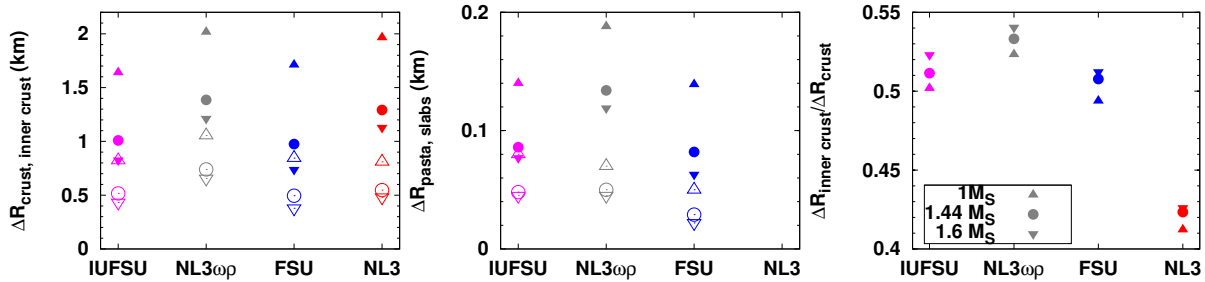} 
  \caption{(Colors online) {Left: thickness of the crust (full symbols) and of the inner crust (empty symbols).
      Middle: thickness of the non-droplet pasta (full symbols) phase and of the slab (empty symbols) phase.
      Right: fraction of the crust occupied by the inner crust.}
}
\label{crust}
\end{figure*}

\begin{figure*}[thb]
\begin{minipage}[t]{0.7\linewidth}
\centering
  \begin{tabular}{cc}
  \includegraphics[angle=0,width=0.5\linewidth]{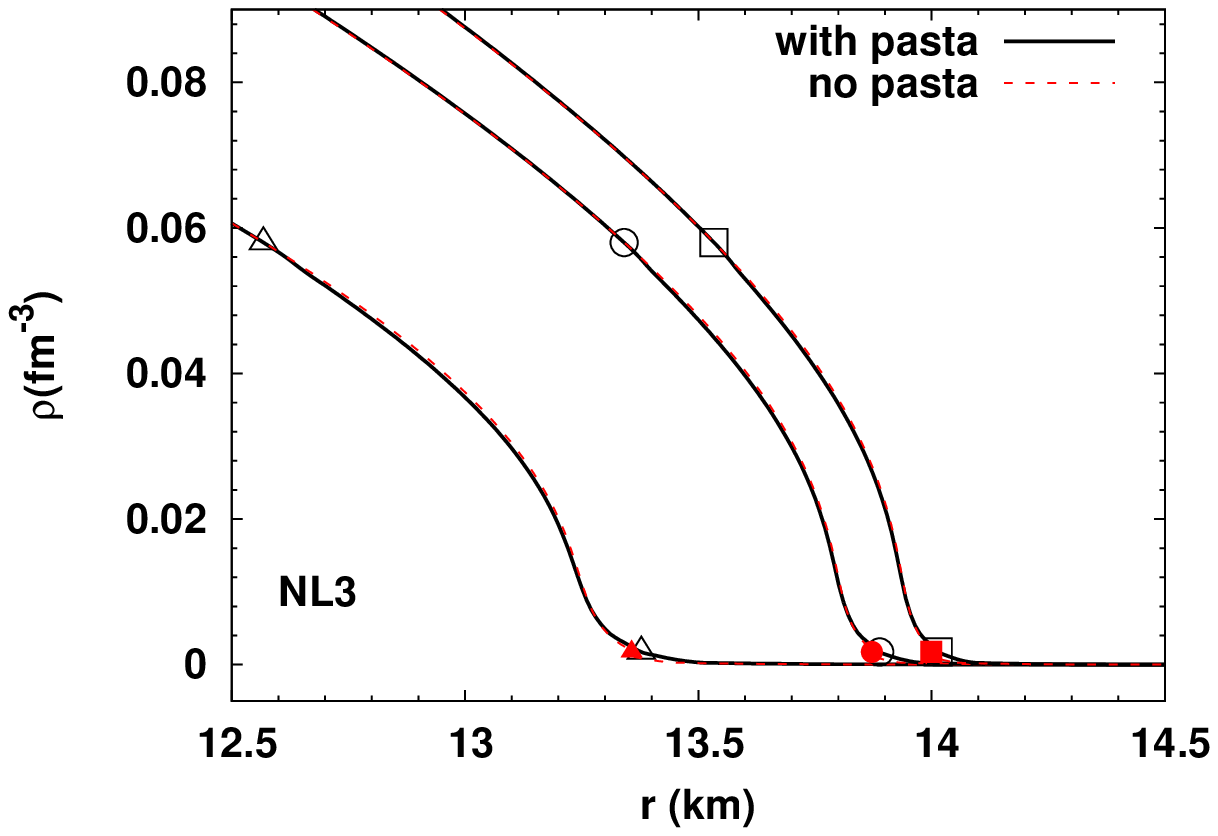}& 
  \hspace{-0.8cm}  \includegraphics[angle=0,width=0.5\linewidth]{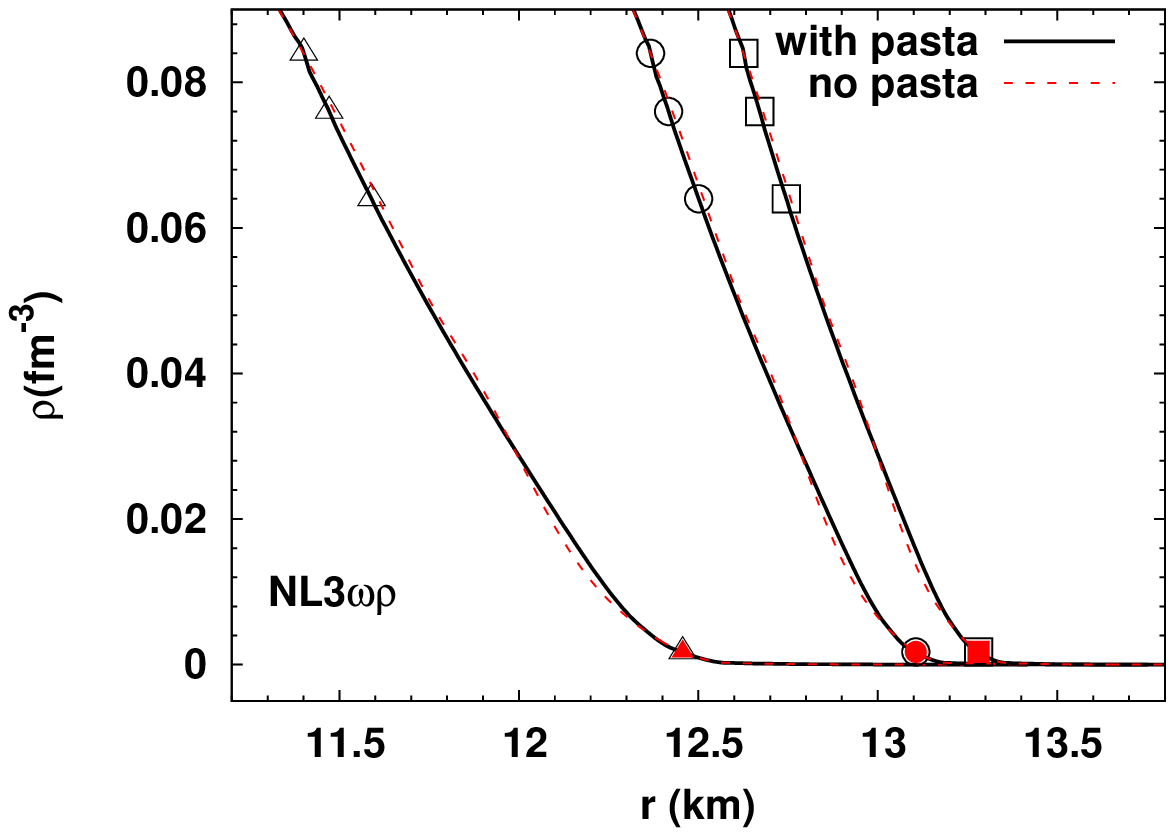}\\ 
  \includegraphics[angle=0,width=0.5\linewidth]{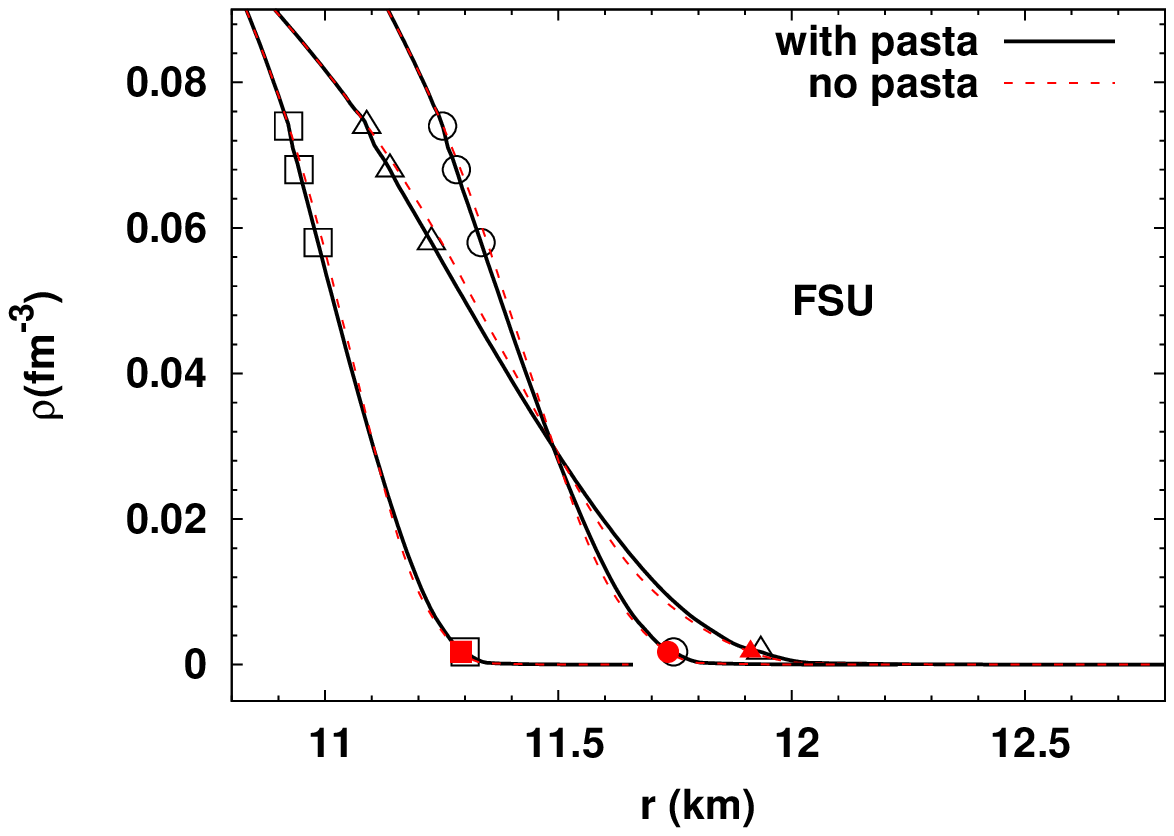}& 
   \hspace{-0.8cm}  \includegraphics[angle=0,width=0.5\linewidth]{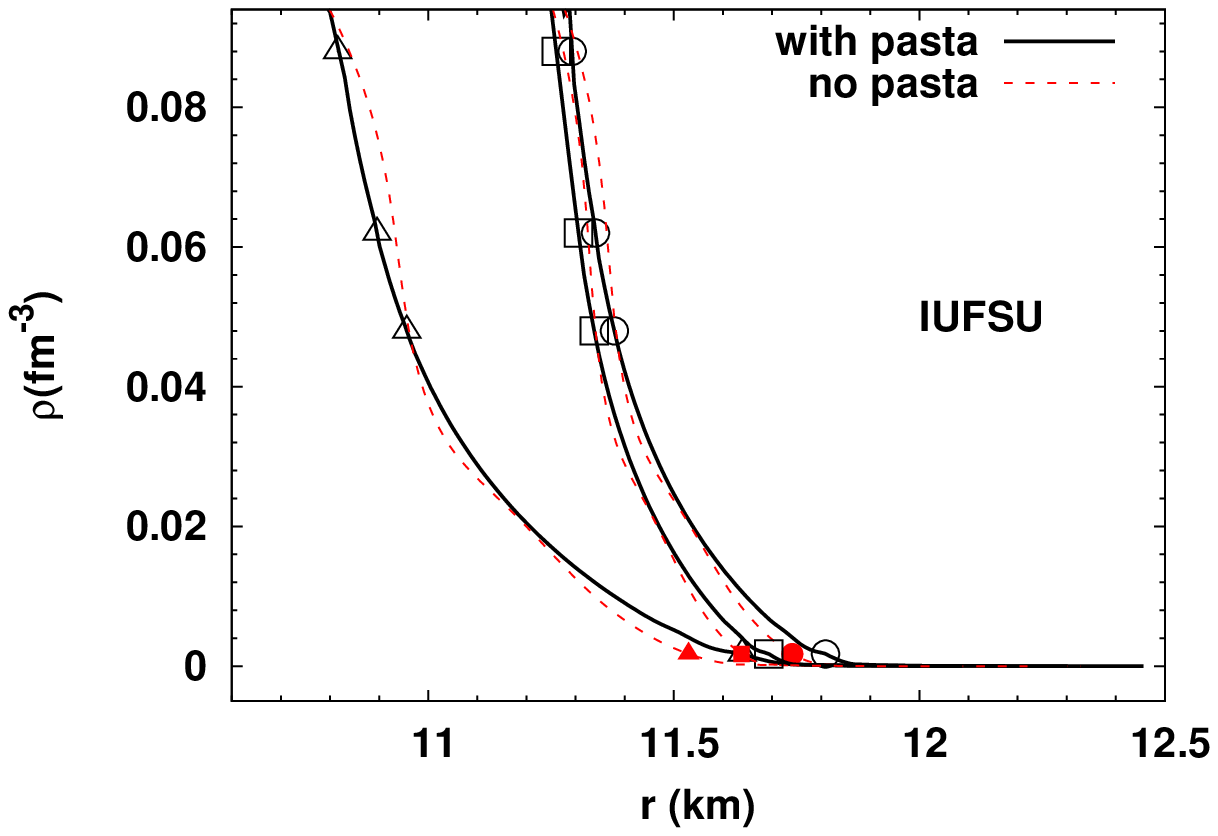}\\ 
  \end{tabular}
\end{minipage}
\begin{minipage}[l]{0.3\linewidth}
 \caption{(Colors online) Crust of neutron stars with a mass equal to
    1.0, 1.44 and 1.6 $M_\odot$ 
    obtained for different nuclear models (solid lines).
    The empty symbols stay at the
    BPS-droplet, droplet-rod, rod-slab, slab-homogeneous transition
    for 1.0 $M_\odot$ (circle),  1.44 $M_\odot$ (triangle),  1.6 $M_\odot$ (square).
    {For comparison, the profiles obtained without pasta (dashed) are also shown. 
      The full symbols denote the BPS-homogeneous matter transition 
      in this case.}
\label{tov}
}
\end{minipage}
\end{figure*}


The extension of the
total crust is mainly defined by the incompressibility of the EOS (cfr Table~\ref{tab:JL}).
 However, the fraction of the crust occupied
 by the inner crust is of the order of 50$\%$ or less and  depends on
 the symmetry energy. This quantity increases with
 the star mass, resulting in a
 difference of $\sim + 5\%$ between stars of mass 1.6 and 1.0 $M_\odot$.

The strong differences existing  between NL3 and NL3$\omega\rho$ allow us
to identify the effect of the symmetry energy, since these two models
only differ {with respect to} the density dependence  of the symmetry energy.  It has
already been shown that there is an anti-correlation between the
crust-core transition density and the slope $L$ at saturation
\cite{vidana09,ducoin11} and, therefore, it could be expected a
larger 
crust for  NL3$\omega\rho$. We
point out, however, that this correlation does not exist with the
total crust thickness, but only with the inner crust.
 In a 1.0 $M_\odot$ star the non-droplet pasta
extension is {smaller than} 
200 m. Generally the stars with
smaller mass have smaller relative pasta phases. The slab fraction
corresponds to $\sim35\%$ of the total pasta phase  for all the models, 
apart from IU-FSU, where it is almost $60\%$. 
The different behavior of IU-FSU is mainly due to the small value of
the symmetry energy slope at
subsaturation densities, which affects the surface tension giving quite high
surface tension for different proton fractions, see \cite{grill12}. A large
surface tension favors the slab geometry with respect to the rod geometry. 
On the contrary, a smaller surface tension favors the formation of droplets, clusters
with the largest surface for the same volume, in a larger density range.

In Fig. \ref{tov} we have plotted the last $\sim 2$ km of the star 
profile closer to the surface. A larger mass corresponds  
to a steeper profile as expected, due to the larger gravitational force.
For NL3 and NL3$\omega\rho$, both with a large incompressibility,
the star with the                                                                                                                
larger mass has the  inner crust at a larger distance from the                                                                    
center.  In the case of FSU and IU-FSU there is a larger concentration                                                                     
of mass at the center because the EOS is softer,  and the crust is                                                                                           
 pushed  more strongly {towards} the center of the star: this explains why
 for IU-FSU the profiles of the 1.44 and 1.6
 $M_\odot$ stars are almost
 coincident, and for FSU the profiles of the 1.0 and 1.44 $M_\odot$ {stars}
 cross, while the {crust} 
of the 1.6 $M_\odot$ {one} has the {smallest} distance to
 the star center.  One interesting conclusion is that  taking into
 account  the  correct
 description of the inner crust in the total stellar EOS is more
 important for the softer EOS and with smaller slopes $L$. However, on the whole,
 using the BPS  EOS for the  outer crust and an EOS of homogeneous
 stellar matter for the inner crust and core gives good results { for
 the stellar profiles.}

In \cite{newton2011} the effect of the nuclear pasta on the crustal shear
phenomena was studied. In particular, two limits have been considered, namely the
pasta as an elastic solid and as a liquid. In the first case the shear modulus
is calculated { from the  crust-core transition} while in the second case at the
transition {from the} droplet to the pasta phase. For models with no pasta phase, as
NL3, there is no difference between these two pictures. However, models with a
symmetry energy slope $L$ below 80 MeV have a pasta phase and the ratio shear
modulus to pressure may be as high as two times larger if the first picture is
considered for $L=40$ MeV. 

\section{Symmetry energy and the strangeness content of a neutron star}
For stellar matter with hyperonic degrees of freedom, as the one
described in this section,  the electromagnetic-field is switched
off, the sum over
nucleons in Eq. (\ref{lagdelta}) is replaced by a sum over the octet of lightest baryons 
(n,\, p,\, $\Lambda,$ $\Sigma^-,$  $\Sigma^0,$ $\Sigma^+,$ $ \Xi^-,$ $\Xi^0$), and the
couplings of the mesons to the baryons are baryon dependent.
Due to the Pauli principle the nucleon Fermi energy increases
and, if the Fermi energy of nucleons becomes larger than the hyperon masses,
energy and pressure are lowered by conversion of some nucleons into
hyperons. This softens the equation of state and has some direct consequences
on the properties of compact stars: maximum star
masses become smaller and neutrino fractions in neutrino trapped
matter are larger.

To fix the model, we need to define the couplings $g_{ij}$, where $i$ is any meson 
and $j$ any baryon of the octet.
For the nucleonic sector, we use the IU-FSU \cite{iufsu} and TM1 \cite{tm1} parametrizations. 
{The latter is a parametrization that satisfies the heavy-ion flow
constraints for symmetric matter at 2-3$\rho_0$ \cite{danielewicz}.
To better understand the effect of the symmetry energy  on the strangeness content, the mass and radius of the stars, 
we will consider a modified version of IU-FSU with $\Lambda_\omega$ as a free parameter. Analogously, for TM1, we will discuss
a modified version obtained including a non-linear $\omega-\rho$ term (TM1$\omega\rho$) that will allow to change
the density dependence of the symmetry energy,  as presented in \cite{providencia13}}, {when $\Lambda_{\omega}$ runs from 0 (TM1) to 0.03. For the slope of the symmetry energy at saturation density we have the following values: $\Lambda_{\omega} = 0$ ($L = 110$~MeV), $\Lambda_{\omega} = 0.01$ ($L = 80$~MeV), $\Lambda_{\omega} = 0.02$ ($L = 70$~MeV) and $\Lambda_{\omega} = 0.03$ ($L = 55$~MeV). }

For the hyperons, {we consider two different sets of hyperon-meson couplings, that we name A and B.}
Within the coupling
set A \cite{chiap09} the $\omega$ and $\rho$ meson-hyperon coupling constants are obtained 
using the SU(6) symmetry:
$ \frac{1}{2} g_{\omega \Lambda} = \frac{1}{2} g_{\omega \Sigma} = g_{\omega \Xi} = \frac{1}{3} g_{\omega N}$, 
$\frac{1}{2} g_{\rho \Sigma} = g_{\rho \Xi} = g_{\rho N},$ $ g_{\rho \Lambda} = 0$,
where $N$ means `nucleon' $(g_{iN} \equiv g_{i})$. The coupling constants 
$g_{\sigma Y} $ of the hyperons with the scalar 
meson $\sigma$ are constrained
by {choosing} the hypernuclear potentials in nuclear matter to be consistent with hypernuclear 
data \cite{hyp1}. 
{Namely, we impose} (see  Ref. \cite{hyp1})
$U_{\Lambda} = -28~{\rm MeV}, \; U_{\Sigma} = 30~{\rm MeV}\,,\, 
U_{\Xi} = -18~{\rm MeV}$, 
{being} 
$U_j = x_{\omega j} \, U_{\omega} - x_{\sigma j} \,
U_{\sigma} $
where $x_{ij} \equiv g_{ij}/g_{i}$, $U_{\omega} \equiv g_{\omega} \omega_0$~and 
$U_{\sigma} \equiv g_{\sigma} \sigma_0$~are the nuclear potentials for symmetric 
nuclear matter at saturation.

\begin{figure}[tbh]
\begin{center}
\includegraphics[width =.7\linewidth]{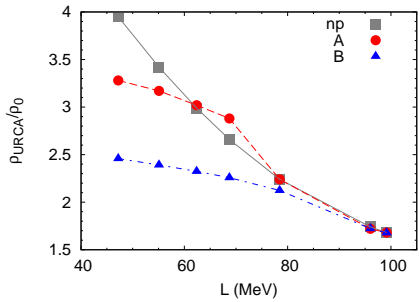} 
\end{center}
\caption{(Color online) 
Onset density of the direct Urca process in stellar matter for nucleonic matter (gray squares), hyperon
coupling set A (red  circles) and  hyperon
coupling set B (blue triangles) for IU-FSU and modified versions with  $L=47-$99 MeV. 
 }
\label{urca}
\end{figure}

In order to show how results are
sensitive to the hyperon couplings we consider 
 {a quite different} 
set of couplings
 proposed in \cite{gm1},  set B,  with  $x_{\sigma Y}$=0.8 and equal for all the hyperons. The
fraction $x_{\omega Y}$  is determined using  $U_Y = -28~{\rm MeV}$ 
for all the hyperons. For the hyperon-$\rho$-meson coupling we consider
$x_{\rho Y}=x_{\sigma Y}$. This choice { implies} 
that the interaction of all hyperons in
symmetric nuclear matter is attractive, and is restricted by acceptable
maximum mass star configurations.

The onset density of the nucleon Direct Urca (DU) process is
plotted as a function of the slope $L$ for the IU-FSU and
modified versions in Fig. \ref{urca}.
The effect of the symmetry energy and the hyperon interaction on the DU onset
density can be summarized as follows
a) for non-strange matter the larger the slope $L$ the smaller the
neutron-proton asymmetry above the saturation density and, therefore, the
smaller the DU onset density;
b) generally, for a low value of $L$ the presence of
hyperons decreases the onset density. This is always true if the hyperon onset
occurs with a negatively charged hyperon because the proton fraction
increases. However, if the hyperon onset occurs with a neutral hyperon, both the proton and the neutron
fractions decrease and it is the net effect that defines the behavior.

In  Fig. \ref{strange} 
the strangeness fraction ($f_s=\sum_j|s_j|n_j/3 n_B$
with $s_j$ and $n_j$ the strangeness,
partial density of the baryon $j$and $n_B$ the total baryonic density) for
IU-FSU  and the modified
IU-FSU model with $L=99$ MeV is plotted as a function of density. A smaller
symmetry energy slope hinders the formation of hyperons because it  gives
rise to a softer EoS.  The conditions for the onset of
hyperons depend on the charge of the hyperon and on the hyperon interaction: $\Lambda$ is the first hyperon
to  appear with set A and  occurs at larger densities for a smaller slope
$L$, on the contrary, with set B, $\Sigma^-$ will occur first and at
smaller densities  for smaller values of $L$ \cite{rafael11}.

\begin{figure}[tbh]
\begin{center}
\includegraphics[width =.82\linewidth]{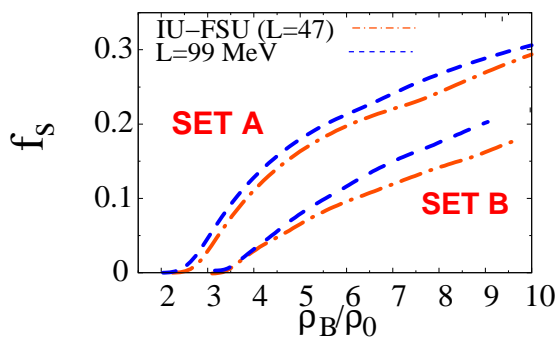} 
\end{center}
\caption{(Color online)
 Strangeness fraction  with the hyperon-meson coupling
  sets A and B for IU-FSU ($L=47$ MeV) and a modified
  version with $L=99$ MeV. }
\label{strange}
\end{figure}

\begin{figure}[tbh]
\begin{center}
\includegraphics[width =1.\linewidth]{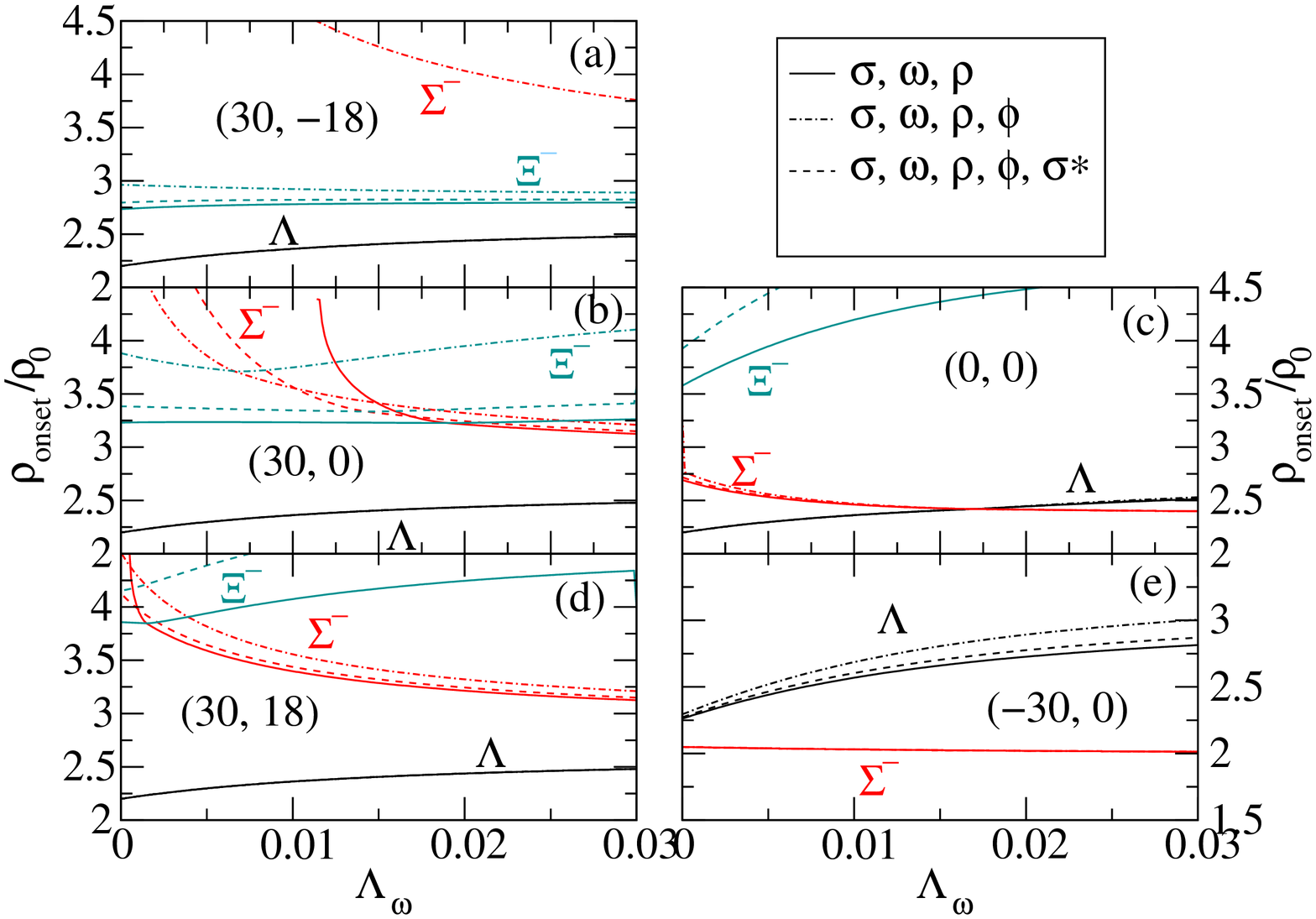} 
\end{center}
\caption{(Color online) Onset of the $\Lambda$, $\Sigma^-$ and $\Xi^-$
  hyperons as a function of the parameter $\Lambda_{\omega}$, with
  $U_\Lambda=-28$ MeV and several values of $U_\Sigma$ and $U_\Xi$
  in symmetric nuclear matter at saturation. The limiting values of
  $\Lambda_{\omega}$, 0 and 0.03, correspond to L=110 MeV and 55 MeV respectively.
  The
  pair of values given in each graph refers to $(U_\Sigma,U_\Xi)$. 
  {Results obtained for the modified TM1 model and extended set A.}
}
\label{onset}
\end{figure}

{From now on, we focus on the TM1 parametrization, and, 
for the hyperon couplings, on the set A.
However, we allow some variations in it. In fact,} while the binding of the
$\Lambda$ in symmetric nuclear matter is well settled experimentally, 
the  binding values of the $\Sigma^-$ and $\Xi^-$
still have a lot of uncertainties \cite{gal2010}. {We, therefore, 
enlarge the set A by allowing $U_{\Xi} =-18,0,+18$ MeV and 
$U_{\Sigma} = -30,0,30$ MeV.}
Finally, we also consider the inclusion of the strange mesons
$\sigma^*,\,\phi$ { to take into account the YY interactions}. 
According to recent experimental    
$\Lambda-\Lambda$-hypernuclear data, the $\Lambda-\Lambda$ interaction is
 only weakly attractive \cite {gal2011}. 
The effect of the small attractiveness of the 
hyperon-hyperon coupling will be considered by choosing 
a) a weak $g_{\sigma^* Y}$ coupling according to Eq. 5 of \cite{cavagnoli08}; b) the extreme value $g_{\sigma^* Y}=0$.
{In both cases, the $\phi$ meson couplings are fixed according to 
$2 g_{\phi \Lambda}= 2 g_{\phi \Sigma} =g_{\phi \Xi}=-\frac{2 \sqrt{2}}{3} g_{\omega N}$.}

\begin{figure}[tbh]
\begin{center}
\includegraphics[width =.7\linewidth]{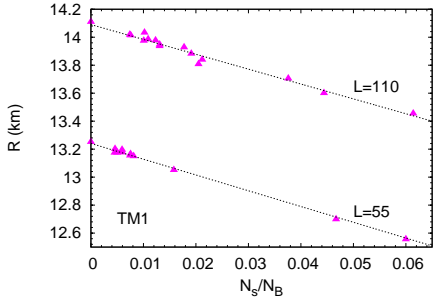} 
\end{center}
\caption{(Color online) Radius of a star with mass $1.67\, M_\odot$ as a function of the
  strangeness content, for TM1 with $L=110$ MeV (top line with slope -11.27
  $\pm 4\%$ km) and TM1 with
  $\omega\rho$ term and $L=55$ MeV (bottom line with slope -10.62 $\pm 1\%$
  km).
  {For a given L, each point corresponds to a choice for the hyperon couplings 
    within the extended set A.}
}
\label{svr}
\end{figure}

{ As mentioned above}, the symmetry energy also affects the onset of hyperons. 
In Fig.~\ref{onset} 
it is shown that the different hyperons are
affected in a different way by the symmetry energy. In this figure we
plot the onset of the $\Lambda$, $\Sigma^-$ and $\Xi^-$
as a function of the coupling $\Lambda_\omega$, {where}
$\Lambda_\omega=0$ (0.03) corresponds
to $L=110 (55)$ MeV. It is
seen that the onset of $\Sigma^-$ always decreases with the decrease of $L$,
due to its larger isospin. On the other hand, the onset of $\Lambda$ occurs at
larger densities. The $\Xi^-$ is never  the first hyperon to appear due to its
large mass, but, according to the attractiveness of its potential in nuclear
matter, it can appear as the second hyperon. If the repulsiveness of the
$\Sigma^-$ in nuclear matter is confirmed  we may expect that the $\Lambda$ is
the first hyperon to set on and, therefore, with a smaller slope $L$ the onset
of strangeness occurs at larger densities. However, if the optical potential
of the $\Sigma^-$ in nuclear matter is only slightly repulsive there may be a
competition between the onset of $\Lambda$ and $\Sigma^-$ depending on the
$L$, with smaller values of $L$ favoring the $\Sigma^-$ hyperon (see top
figure of the right column).

As discussed in \cite{rafael11,panda12}, a smaller slope
$L$ implies a softer increase of the strangeness fraction with density.
However, {once} the central density of these stars is larger, it is important
to study { their} total hyperon content. This will be done by calculating for each star the total
strangeness number
$$N_S=4\pi\int_0^R \frac{\rho_s\, r^2}{\sqrt{1-2m(r)/r}} dr,$$
where $m(r)$ is the mass inside the radius $r$.

In Fig. \ref{svr} we plot the radius of a star with a mass 1.67$M_\odot$
similar to the mass of 
the pulsar PSR J1903+0327 ($1.67\pm 0.02 M_\odot$) \cite{j1903} as a function of
its strangeness content. The largest strangeness fractions were obtained
considering an attractive potential for the $\Sigma^-$ meson. It is
interesting to notice that two almost parallel straight lines are obtained:
for $L=110$ MeV the  slope  is -11.27  $\pm 4\%$ km and
for $L=55$ MeV  the slope is  -10.62 $\pm 1\%$ km. The straight
lines cross the vertical axis for a nucleonic star with no hyperons.
The slope is almost independent of $L$.

\begin{figure*}[t]
\centering
\includegraphics[width=0.95\linewidth,angle=0]{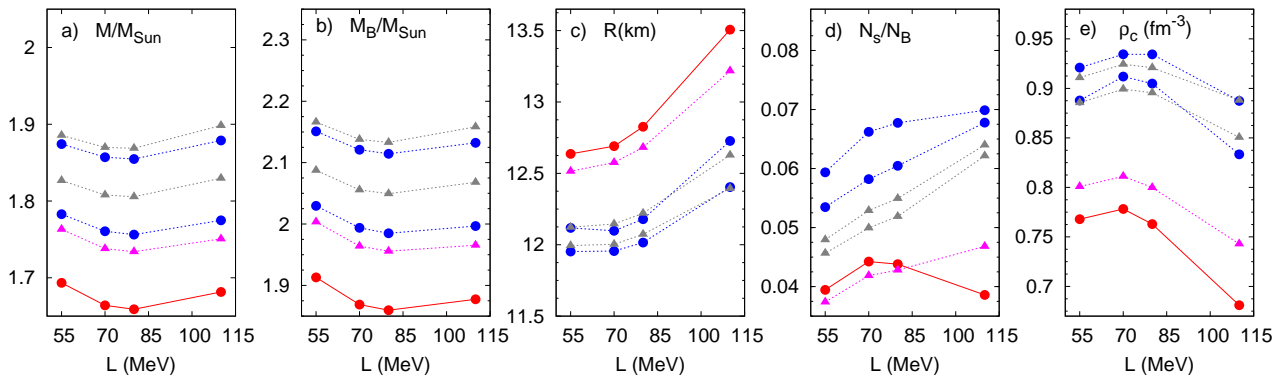} 
\caption{(Color online) 
Properties of maximum mass stars obtained with $L=55, 70, 80$ and 110 MeV,
for TM1 {and extended set A}:  $U_\Xi=+18$ MeV (triangles)  without (pink) and  with (grey) $YY$ interaction; 
$U_\Xi=-18$ MeV  (circles)  without (red) and with (blue) $YY$ interaction. 
{For the YY interaction, two choices are considered, as discussed in the text.
For all the curves, $U_\Lambda=-28$ MeV and $U_\Sigma=30$ MeV.
The panels contain:}
 a) gravitational mass, b) baryonic mass, c) star radius, d) strangeness content and e) central density.}
\label{MR1}
\end{figure*}
{ Finally, in \Fig{MR1} we show some properties of maximum mass stars.}
Some general conclusions may be drawn with respect to the strangeness
content:
a) the maximum star  mass changes with $L$, stars with an intermediate $L$ have
the smallest masses and, generally, have the largest central densities (see
panels a) and e)).  There
are two competing factors that define this behavior: on one hand a larger $L$
corresponds to a harder EOS because the symmetry energy increases faster with
the density, on the other hand a larger $L$ favors larger strangeness
fractions which softens the EOS. The first one gives rise to smaller central
densities and larger radii, while the second one leads to the opposite; 
b) the strangeness content depends on the hyperon interaction, and, 
in particular, on the $\Xi$ potential in the present study. If $U_\Xi=+18$ MeV (triangles)
the masses are larger and the strangeness fractions generally smaller;
c) {the inclusion of} the strange mesons gives rise to more
massive stars that may have larger
strangeness contents. In this case the strangeness content is always smaller
for a smaller slope $L$, and its maximum value is of the order
0.04-0.05 according to the hyperon interaction if $L=55$ MeV. 
The upper limit can reach 0.07-0.08 if $L=110$ MeV,
Fig.~\ref{MR1}d). Larger fractions may be obtained if the $U_\Sigma$ is
considered attractive.

\section{Conclusions}
We have studied the effect of the density dependence of the symmetry energy
on several properties of neutron stars. In particular, we have discussed the
{properties of the crust-core transition, the pasta phase,} and a possible
existing  competition between the effects of  the
symmetry energy and exotic degrees of freedom in the EOS.

The problems have been investigated within different nuclear matter approaches,
namely {the BHF one, several Skyrme forces, RMF models and a generalized liquid drop
model.} 

{First, we have analyzed the correlations of the slope parameter $L$ 
with the core-crust transition  from 
homogeneous to clusterized matter in neutron stars, using a simplified definition of this transition, namely the crossing between the line of beta equilibrium and the thermodynamical spinodal.
 It was shown that the
core-crust transition density  and proton fraction appear clearly correlated with $L$
\cite{ducoin10,ducoin11}. On the other hand no clear correlation was observed
between  $L$ and the transition pressure.

We have shown that the determination of the core-crust transition
by the crossing between the dynamical spinodal and the $\beta$-equilibrium  
corresponds to a realistic approximation, with results very similar to the TF prediction. 
It takes place at a density lower than expected if the thermodynamical spinodal 
approach is applied, however,  it was confirmed that the correlations obtained within the  thermodynamical
approach are still valid using the dynamical spinodal approach.

We have verified that the  predictivity of the transition pressure  is considerably
improved in terms of selected pairs of coefficients. In particular,
a strong correlation appears between  the transition pressure and a combination of the symmetry 
energy slope and curvature parameters at the same reference density, $\rho=0.1$ fm$^{-3}$.
This correlation indicates that the relation between nuclear
observables and the liquid drop model coefficients should be investigated at subsaturation densities.}

{ In the second part,} we have studied the inner-crust properties of neutron
stars within a self-consistent Thomas-Fermi approach developed
in Refs. \cite{pasta1,pasta2} for relativistic nuclear models, and 
the coexisting-phases
method \cite{pasta3,pasta1}. Several relativistic nuclear models have
been used, with
 nonlinear meson terms and constant couplings, or
with density-dependent coupling constants.

The properties of the models used are reflected in the cluster
structure. It was seen that a small symmetry energy slope $L$
gives rise to larger cells, with a larger proton and neutron
number, while the opposite occurs for a large $L$. 
In particular, it was shown that the NL3 model, with a very large symmetry
energy and slope, and the IU-FSU one, with a quite small slope, have
very different behaviors. NL3 does not present non-droplet pasta
phases in the inner crust of $\beta$-equilibrium matter, and predicts
the smallest proton and neutron numbers and Wigner-Seitz
radii in almost all the inner-crust range of densities.
On the contrary, IU-FSU predicts a quite low density for the onset
of the non-droplet pasta phase, the largest crust-core transition density,
and the largest clusters.

All models, except NL3, predict the existence of  {\it lasagna}-like structures that may
have an important contribution to the specific heat of the
crust \cite{luc2011}.

The effect of the inner crust EOS on the neutron star profile was also
analysed. It was verified that a smaller slope gives rise to a steeper crust {density profile} 
and a larger inner crust with respect to the total crust. It may also enhance
the slab phase extension as observed in IU-FSU.

{Finally, in the last part we discussed} the joint effect of strangeness and the symmetry energy on
some properties of the neutron stars, such as the hyperon content, DU and radius.

 It was  shown that the smaller the slope $L$, the
larger the onset density of the DU process.
  However,
the DU onset also depends on the hyperon content and the
hyperon-meson couplings. The DU may be hindered or favored according to a balance
between the neutron and proton reductions if a neutral hyperon sets on
first. Negatively charged hyperons favor the DU onset due to a decrease of the neutron fraction and an
increase of the proton fraction.

It was also shown that, for a star with a fixed mass, the radius
of the star decreases linearly with the increase of the total
strangeness content. In particular, a 1 km decrease of the radius
of a 1.67 $M_\odot$  star may be explained if the slope of the symmetry
energy decreases from 110 to 55 MeV or the strangeness to
baryon fraction increases from zero to  0.09.

A softer symmetry energy corresponds
to a slower increase of the hyperon fraction with
density. However, the onset of strangeness depends on
the charge of the hyperons. Negatively charged hyperons set
on at smaller densities while neutral hyperons appear at larger
densities for smaller values of the slope. 

If a repulsive  hyperon-hyperon
interaction is considered, although a larger slope $L$ gives
rise to a larger strangeness content, the extra repulsion between
hyperons  compensates 
the extra hyperon fraction and the effect of the symmetry
energy is almost not seen on the central density of the
maximum mass configuration. 

We conclude that some
star properties are affected in a similar way by the density
dependence of the symmetry energy and the hyperon content
of the star. 
To disentangle these two effects it is essential to
have a good knowledge of the EOS at suprasaturation densities.
There is still lack of information about the nucleonic EOS at
supra-saturation densities as well as on the hyperon interactions in nuclear matter 
that may allow {for} an unambiguous  answer to  whether the mass of the pulsars
J1614-2230 \cite{demorest} or J0348+0432 \cite{j0348} could rule out
exotic degrees of freedom from the interior of compact stars.

{The symmetry energy density dependence
  and its slope have been topics of intense investigation in the
  latest years. The work we have just presented in this paper is
  intrinsically related to other topics also discussed in this special
  volume. The search for constraints to  the huge variety of equations
  of states used to describe neutron star matter involves 
  astrophysical observations, heavy ion collision data, nuclear
  reactions \cite{lattimer,tsang},  nuclear structure,  
 bulk matter empirical values \cite{jirina} and
finite nuclei properties, as the neutron skin thickness
\cite{fattoyev,rocamaza}.
On the other hand, the strangeness content of different equations of
state, shown above to be related to the symmetry energy, has important
consequences on both the liquid-gas phase transition and the
transition at high densities \cite{francesca}. Hence, further
investigation towards a better understanding of the symmetry energy 
density dependence is still required. }

{\bf Acknowledgements:}

This work was carried out within the R\&DT projects PTDC/FIS/113292/2009
and CERN/FP/123608/2011, developed under the scope of a QREN initiative,
UE/FEDER financing, through the COMPETE programme and  by the NEW COMPSTAR, a COST initiative.


\end{document}